\pdfoutput=1
\documentclass[aps,pra,10pt,twocolumn,showpacs,superscriptaddress,nofootinbib]{revtex4-1}
\usepackage{amsmath}
\usepackage{latexsym}
\usepackage{amssymb}
\usepackage{bm}
\usepackage{color}
\usepackage{amsmath}
\usepackage[usenames,dvipsnames,svgnames]{pstricks}
\usepackage{epsfig}
\usepackage{pst-grad} % For gradients
\usepackage{pst-plot} % For axes
\usepackage{newlfont}
\usepackage{amsfonts}
\usepackage{amsthm}
\usepackage{graphicx}
\usepackage[breaklinks]{hyperref}
\usepackage{tikz}

\definecolor{lime}{HTML}{A6CE39}
\DeclareRobustCommand{\orcidicon}{%
	\begin{tikzpicture}
	\draw[lime, fill=lime] (0,0) 
	circle [radius=0.16] 
	node[white] {{\fontfamily{qag}\selectfont \tiny ID}};
	\draw[white, fill=white] (-0.0625,0.095) 
	circle [radius=0.007];
	\end{tikzpicture}
	\hspace{-2mm}
}

\foreach \x in {A, ..., Z}{%
	\expandafter\xdef\csname orcid\x\endcsname{\noexpand\href{https://orcid.org/\csname orcidauthor\x\endcsname}{\noexpand\orcidicon}}
}

\begin{document}
\title{\textcolor{black}{Semi-relativistic} (e,2e) study with twisted electron beam on Cu and Ag}
\affiliation{Department of Physics, Birla Institute of Technology and Science, Pilani, Pilani Campus, Pilani,  Rajasthan, 333031, India}
\affiliation{Univ Rennes, CNRS, IPR (Institut de Physique de Rennes) - UMR 6251, F-35000 Rennes, France}
\author{Aditi Mandal$^{1}$ \orcidA{}, Nikita Dhankhar$^{1}$ \orcidB{}, Didier S\'{e}billeau$^{2}$ \orcidC{} and Rakesh Choubisa$^{1}$ \orcidD{}}
\email{rchoubisa@pilani.bits-pilani.ac.in}


% \date{\today}

\begin{abstract}
In this communication, we report our calculations of Triple Differential Cross-Section (TDCS) for the relativistic (e,2e) process with a twisted electron beam on Cu and Ag atomic targets in coplanar asymmetric geometry mode. The theoretical formalism has been developed in the first Born approximation (FBA) in which we use the Dirac plane wave as well as the twisted electron wave for the incident electron beam to study the effect of various parameters of the twisted electron beam on the (e,2e) process. We use  Dirac plane wave, semi relativistic Coulomb wave and Darwin wave function for the scattered, ejected and K-shell electron respectively. We compare the angular profiles of the TDCS  of the twisted electron impact (e,2e) process with that of the plane wave. We segregate the TDCS for charge-charge interaction and current-current interaction with their interference term and study the effect of \textcolor{black}{different parameters} of the twisted electron beam on them. The study is also extended to the macroscopic Cu and Ag targets to further investigate the effect of the opening angle of the twisted electron beam on TDCS. The spin asymmetry in TDCS caused by polarized incident electron beam is also studied to elucidate the effects of the twisted electron beam on the (e,2e) process.

\end{abstract}

\maketitle 

%%%%%%%%%%%%%%%%%%%%%%%%%%%%%%%%%%%%%%%%%%%%%%%%%%%%%%%%%%%%%%%%%%%%%%%%%%%%%%
\section{Introduction}\label{sec1}
The coincidence (e,2e) studies on \textcolor{black}{numerous} atomic and molecular \textcolor{black}{targets} have been explored for the last five decades for the impact energy ranging from low energy to relativistic energy. (e,2e) field has a long history in atomic and molecular physics~\cite{a49}. Originally derived for the (p,2p) spectroscopy in nuclear physics~\cite{a2}, where p represents a proton, it was proposed in 1966 by Smirnov and co-workers \textcolor{black}{to the (e,2e) processes on atoms} for the investigation of atomic wave functions~\cite{a3,a4}. Since then, it has enjoyed a widespread application, such as electron momentum spectroscopy \textcolor{black}{in condensed matter physics}~\cite{a5}. In many branches of physics, such as astrophysics and plasma physics, there has been considerable interest in the study of ionization processes by a charged particle. The electron-impact single ionization called the (e,2e) process, has thus become a powerful tool for investigation of the dynamics of the ionization process~\cite{a6}. The coincidence cross-sections \textcolor{black}{for the (e,2e) process}, here defined as Triple Differential Cross Sections (TDCS), depend\textcolor{black}{s} on the momenta of two outgoing electrons (scattered and the ejected electron\textcolor{black}{s}). Since the first (e,2e) measurements reported independently in the late 1960s by Ehrhardt et al.~\cite{a7} and Amaldi et al.~\cite{a8}, experimental and theoretical activities in the non-relativistic energy scale have been intense (e.g. see~\cite{a9,a10,a11}). The (e,2e) \textcolor{black}{field is still an  active field in the current time with more focus} on molecular targets. A short time ago, various variants of theoretical models\textcolor{black}{,} based on \textcolor{black}{the} 3-Coulomb approach by Brauner et al.~\cite{a46}\textcolor{black}{,} have been used to study \textcolor{black}{(e,2e) process on} argon atom and $H_{2}O$, $CH_{4}$ and $NH_{3}$ \textcolor{black}{molecules}~\cite{a12}. In all the above-mentioned works, the electron impact energy lies in an energy range typically between 10 eV and 10 keV wherein the spin aspects do not play a major role except in anti symmetrization of the wave-function of involving electrons. Work in the relativistic energy regime began in 1982 with the absolute (e,2e) experiments of Schule and Nakel~\cite{a13} at an incident energy of 500 keV on the K shell of the silver atom. These kinematically complete experiments on the inner shell states of high-Z atoms probe the fundamental ionization mechanism in the regime of relativistic energies. Later on, experiments have been performed with transversely polarized electron beams~\cite{a14} in which, apart from the momenta, the spin of \textcolor{black}{the} impinging beam is also resolved. These (e,2e) experiments have entailed the development of new theoretical and computational methods~\cite{a15}.

Recently, there have been new interesting breakthroughs to create electron vortex beam which has the ability to carry Orbital Angular Momentum (OAM) along the propagation direction of the electron beam. It is termed  as ``{\it{twisted}} electron" beam~\cite{a16}. Hence with this, now we would be able to probe multiple sources of perturbation\textcolor{black}{s in a system, for instance\textcolor{black}{,} it can be employed as \textcolor{black}{a} nanoscale probe in magnetic materials}. Initially, the concept of twisted photon beam came into the picture and accordingly researchers have begun to appreciate its implications \textcolor{black}{on} our \textcolor{black}{basic} understanding of the \textcolor{black}{way the} light \textcolor{black}{interacts with the matter for conventional photon beam.} \textcolor{black}{Researchers also realized its} potential for quantum information applications~\cite{a17}. \textcolor{black}{On} similar lines, \textcolor{black}{there have been intense studies on the production and application of the twisted electron beam}~\cite{a18,a19,a50,a51}. \textcolor{black}{The} twisted electron beam is not a plane wave, but \textcolor{black}{it is a} superposition of plane waves with a defined projection of the \textcolor{black}{OAM} onto the propagation axis. This projection, which nowadays can be very high~\cite{a20,a21}, determines the magnitude of the OAM induced magnetic moment. Due to such a huge magnetic dipole moment (as opposed to the plane-wave electron \textcolor{black}{beam}), twisted electron \textcolor{black}{beam} is presently regarded as a valuable tool for studying the magnetic properties of materials at the nanoscale~\cite{a22,a23,a24}.

\textcolor{black}{In the recent past}, the first Born approximation and Dirac's relativistic theory have been applied to explore the Mott scattering of high-energetic twisted electron \textcolor{black}{beam} by atomic and macroscopic targets to study the effects of total angular momentum (TAM) \textcolor{black}{projection}, opening angle and the impact parameter \textcolor{black}{of the twisted electron beam}~\cite{a25}. A generalised Born approximation has been \textcolor{black}{used} to investigate the scattering of the vortex electron \textcolor{black}{beam} by atomic targets~\cite{a26}. Based on the developed theory~\cite{a27,a28,a29,a30}, it has been shown that the number of scattering events in the collisions involving twisted electrons are comparable to that in the standard plane-wave regime. \textcolor{black}{Further, a fully relativistic calculation of \textcolor{black}{the differential cross-section for} the bremsstrahlung, emitted by twisted electrons in the field of bare heavy nuclei, has been done recently by Groshev et. al.~\cite{a54}}. \textcolor{black}{In this communication, we extend such type of study to (e,2e) processes on atoms at relativistic energy}. In the coincidence (e,2e) process, we detect all the participating particles in the continuum state with their momenta fully resolved. Naturally considering the twisted electron \textcolor{black}{beam} in place of the standard plane-wave \textcolor{black}{beam} will extract more information about the (e,2e) processes as available presently. Till now, almost all the (e,2e) activities have been confined to electron beam, which carries linear momentum in its impinging direction except the recent studies by Harris et al.~\cite{a31} \textcolor{black}{and Dhankhar and Choubisa~\cite{a57}} in which (e,2e) processes \textcolor{black}{have} been studied \textcolor{black}{on H atom and $H_{2}$ molecule with twisted electron beam at non-relativistic energy regime}. \textcolor{black}{The relativistic (e,2e) processes on atomic target\textcolor{black}{s} with conventional electron beam have been studied extensively}~\cite{a52,a53}. To the best of our knowledge\textcolor{black}{,} relativistic or semi-relativistic (e,2e) study with twisted electrons has not been explored in the literature, even the theoretical estimation is not explored. The twisted electron carries TAM, consisting of OAM and spin of the electrons, in addition to its linear momentum at relativistic energy ~\cite{a47}. Hence, it will be an interesting task to probe the effects of TAM \textcolor{black}{projection} of the twisted electrons on the (e,2e) processes on atoms\textcolor{black}{,} especially at relativistic energy \textcolor{black}{regime}. Further, one can also probe the effect of twisted electrons on the spin asymmetry in TDCS. The present communication is intended to cover these aspects in a theoretical manner. 

\textcolor{black}{In our theoretical model, we use \textcolor{black}{the} first Born approximation which is a reasonably good approximation for (e,2e) studies on lighter atomic targets. For the conventional plane-wave beam, we use the Dirac plane wave and for the twisted electron beam, we use twisted electron wave function. We describe the scattered, ejected and K-shell electrons by the Dirac plane wave, semi relativistic Coulomb wave and Darwin wave function respectively}. We neglect the exchange-correlation effect arising from the \textcolor{black}{electron cloud of the heavy atomic targets}. In addition to this, we \textcolor{black}{do not} consider the exchange effect between the incident and scattered electrons because here we consider the coplanar asymmetric geometrical mode for our study\textcolor{black}{.} \textcolor{black}{In this geometry, the} energy of the scattered electron ($E_s$) is sufficiently larger than that \textcolor{black}{of the} ejected electron ($E_1$)(the ratio $R=\frac{E_s}{E_1}$ varies from 2.69 to 3.1 \textcolor{black}{in} the \textcolor{black}{paper}). Further, Keller et. al. (1999)~\cite{a32} reported that the exchange effects \textcolor{black}{in (e,2e) process} play a more important role for the heavier target, like Au, than that for the lighter atoms, \textcolor{black}{like Cu and Ag} targets considered here. Generally, FBA is inadequate when the scattering involves many electron atomic targets, although it serves as an important baseline for comparison of more advanced calculations. We hope that the present results may stimulate new types of theoretical and experimental studies on the relativistic (e,2e) processes with twisted electron impact on atoms. Here we present our calculation of TDCS and spin asymmetry for K-shell ionization of Cu and Ag target for twisted electron case. We would like to point out that, in literature, we have better theoretical models for plane wave incidence, for example, relativistic Distorted Wave Born Approximation (rDWBA) ~\cite{a32, a33}, than the \textcolor{black}{first Born approximation (FBA)} presented here. Therefore, the results may not be as accurate as that for the more accurate theoretical model, such as rDWBA, for the twisted electron.

\begin{figure}[ht]
\includegraphics[width = 0.9\columnwidth]{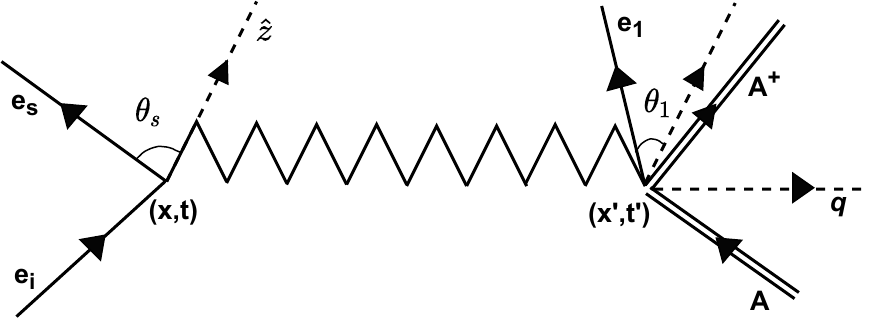}
\caption{Schematic diagram for (e,2e) process on atom (A) by an incident electron $(e_{i})$ in one photon-exchange approximation. The electromagnetic interaction is mediated through a virtual photon along the momentum transfer direction. The incident electron scatters at $\theta_{s}$ angle and the ejected electron ejects at $\theta_{1}$ direction from the incident direction (z-axis) in the scattering plane defined in the xz-plane. The geometry here used is coplanar asymmetric ($E_{s}>E_{1}$).}\label{fig1}
\end{figure}

\textcolor{black}{The paper is organised as follows, followed by the Introduction, we describe our semi relativistic theoretical model for the computations of TDCS and spin asymmetry in TDCS for the (e,2e) processes for both the plane and twisted electron beam in section \ref{sec2}. We discuss our theoretical results of TDCS and spin asymmetry in section \ref{sec3}. Finally, we conclude our findings in section \ref{sec4}}.

\section{Theoretical Formalism}\label{sec2}
We develop our formalism with the following assumptions:

(1) The incident electron emits a virtual photon at $(x,t)$ along the direction of momentum transfer which is absorbed by the atom at ($x^{\prime},t^{\prime}$) (see Fig.~\ref{fig1}). The incident electron-atom interaction is of the first order. Due to this interaction, the incident electron is scattered by an angle $\theta_{s}$ from the incident electron direction.

(2) We also assume that the electromagnetic interaction, via a virtual photon interaction, emits the K-shell bound electron into the continuum state. The ejected electron is ejected in the $\theta_{1}$  direction from the direction of the incident electron. All the electrons are in the same plane (scattering plane). For the incident \textcolor{black}{conventional} electron beam, we use the Dirac plane wave function\textcolor{black}{.} \textcolor{black}{We use} the Bessel wave function for the twisted electron \textcolor{black}{beam}.

(3) We describe Dirac plane wave function, semi relativistic Coulomb wave function and Darwin wave function for the scattered, ejected and K-shell electron respectively. TDCS is computed here in the coplanar asymmetric geometry ($E_{s}>E_{1}$).

\subsection{Plane wave ionization}\label{sec2a}
In the coincidence ionization processes on any target (e.g., atom, ion, cluster, etc.), we calculate the differential cross sections for various kinematic arrangements of the outgoing electrons involved in the ionization process. The ionization of an atomic target by an electron may be framed as: 
\begin{equation}\label{eq1}
e^{-}_{i}(\textcolor{black}{\lambda})+A\longrightarrow A^{+}+e^{-}_{s}+e^{-}_{1}.
\end{equation}

where \textit{i}, \textit{s} and $1$ represent incident, scattered and ejected electron with $A$ being the target. $\lambda$ is the helicity of the incident electron. In the coincident (e,2e) experiment, the momenta of the scattered and ejected electrons in the continuum state are resolved and hence the coincidence cross-section (here TDCS) depends on the energy of either of the two electrons and directions of the electrons. We compute the TDCS corresponding to an (e,2e) process in the first Born approximation as:
\begin{equation}\label{eq2}
\dfrac{d^{3}\sigma(\textcolor{black}{\lambda})}{d\Omega_{s}d\Omega_{1}dE_{s}}= (2\pi)^{4}\frac{k_{s}k_{1}}{k_{i}}\frac{E_{i}E_{s}E_{1}}{c^{6}}\overline{\sum_{\textcolor{black}{ \mu_{b}}}}\sum_{\textcolor{black}{\lambda_{s} \mu_{1}}}|\langle f|\widehat{S}|i\rangle|^{2}. 
\end{equation}

Here for the case of (e,2e) process, $\widehat{S}$ is the S-matrix operator. $\textcolor{black}{\lambda_{s}} $ \textcolor{black}{is the helicity of the scattered electron}. $\textcolor{black}{\mu_{b}} $ and $\textcolor{black}{\mu_{1}} $ are the spin projections of the bound (K-shell) and ejected electrons in the continuum state respectively. $E_{i}$, $E_{s}$, $E_{1}$ and $k_{i}$, $k_{s}$, $k_{1}$ are the on-shell total energies and momenta of the unbound particles. The spin projections are taken with respect to the propagation direction of the incident electron beam (also defined as the z-axis).

Here, TDCS in Eq.~\ref{eq2} is calculated as an average over \textcolor{black}{the} bound electron and a sum over \textcolor{black}{the} final-state spins of the scattered and ejected electrons. The main task here is to calculate the S-matrix element in the following form~\cite{a34}:
\begin{equation}\label{eq3}
\langle f|\widehat{S}|i\rangle = \frac{-1}{c}\int \textbf{A}_{\mu}(\mathbf{r_1})\textbf{J}^{\mu}(\mathbf{r_1})d^{3}r_{1},
\end{equation}
where $\textbf{A}_{\mu}(\mathbf{r_1})$ can be expressed as,
\begin{equation}\label{eq4}
\textbf{A}^{\mu}(\mathbf{r_1})=\frac{4\pi}{(2\pi)^{3}}\dfrac{[\textcolor{black}{\overline{u}_{\mathbf{k}_{s},\lambda}\gamma^{\mu}u_{\mathbf{k}_{i},\lambda}}]}{[q^{2}-(\frac{\Delta E}{c})^{2}]}[{e^{i\textbf{q}\cdot\textbf{r}_{1}}}].
\end{equation}
Here $\mathbf{r_1}$ is the position vector of the ejected electron with respect to the target, $\textbf{q}=\textbf{k}_{i}-\textbf{k}_{s} $ is the momentum transfer, $\Delta E= E_{i}-E_{s}$, $ \gamma^{\mu}$ are Dirac matrices,  $\textcolor{black}{u_{\mathbf{k}_{i},\lambda}}$ and $\textcolor{black}{u_{\mathbf{k}_{s},\lambda}}$ are Dirac spinors. We describe the Dirac spinor for the free electron, characterized by the momentum $\mathbf{k}$ and the helicity $\lambda$, by 
\begin{equation}\label{eq5}
\textcolor{black}{u_{\textbf{k},\lambda}} = 
\begin{pmatrix}
\sqrt{\varepsilon +c^2} \, \,w^{\lambda}(\hat{n})\\ 
\\
2 \lambda \, \sqrt{\varepsilon -c^2} \,\, w^{\lambda}(\hat{n})
\end{pmatrix}.
\end{equation}
We describe $\varepsilon = \sqrt{c^4 + k^2 c^2}$ (in \textcolor{black}{atomic} units) and $\hat{n}= \frac{\mathbf{k}}{k}$ as the total energy and the propagation direction respectively \textcolor{black}{for the free} electron. $w^{\lambda}(\hat{n})$ is the eigenfunction of the Helicity operator $\frac{\hat{\sigma}. \hat{n}}{2}$, $\hat{\sigma}$ is the standard vector of Pauli matrices. The $w^{\lambda}(\hat{n})$ in the Eq.\eqref{eq5} can be constructed for the electron's propagation in any arbitrary direction $\hat{n}= (\sin\theta_p\cos\phi_p, \sin\theta_p\sin\phi_p, \cos\theta_p)$ with respect to the quantization axis (here the propagation direction of the incident electron is taken along z-axis) for the given polar angle $\theta_p$ and azimuthal angle $\phi_p$~\cite{a55,a56}. It can be described as:
\begin{equation}\label{eq6}
w^{\lambda}(\hat{n}) = \sum_{\sigma= - \frac{1}{2}}^{\frac{1}{2}} e^{-i \sigma \phi_p} \, d_{\sigma \lambda}^{\frac{1}{2}} (\theta_p) \, w^{\sigma}(\textcolor{black}{\textbf{e}_z}).
\end{equation}
Here, $ d_{\sigma \lambda}^{\frac{1}{2}} = \delta_{\sigma,\lambda} \, \cos (\frac{\theta_P}{2}) - 2 \, \sigma\,\,  \delta_{\sigma,-\lambda} \, \sin (\frac{\theta_P}{2})$ and $w^{\sigma} (\textcolor{black}{\textbf{e}_z})$ is the standard Pauli's spinor defined as  $w^{\frac{1}{2}} (\textcolor{black}{\textbf{e}_z})= \begin{pmatrix}
1\\ 
0
\end{pmatrix},$ and $w^{-\frac{1}{2}} (\textcolor{black}{\textbf{e}_z})= \begin{pmatrix}
0\\ 
1
\end{pmatrix}.$ On substitution of $ w^{\lambda}(\hat{n})$ in the expression (5), we can express the Dirac spinor as:
\begin{equation}\label{eq7}
\textcolor{black}{u_{\textbf{k}, \lambda}}= \sum_{\sigma= - \frac{1}{2}}^{\frac{1}{2}} \, e^{-i \sigma \phi_p} \, \,  d_{\sigma \lambda}^{\frac{1}{2}} (\theta_p) \times \begin{pmatrix}
\sqrt{\varepsilon +c^2} \, \,w^{\sigma}(\textcolor{black}{\textbf{e}_z})\\ 
\\
2 \lambda \, \sqrt{\varepsilon -c^2} \,\, w^{\sigma}(\textcolor{black}{\textbf{e}_z})
\end{pmatrix}.
\end{equation}
The quantum number $+\frac{1}{2}$ and $-\frac{1}{2}$ values of $\lambda$ are helicity of electron which represent\textcolor{black}{s} right-handed and left-handed polarization respectively of the electrons (projection of spin along beam direction). 
\\
We describe the atomic transition four-current density for the electron transition from the K shell to the continuum state by $\textcolor{black}{\mathbf{J}^{\mu}}(\mathbf{r_{1}}) $ in the following form:
\begin{equation}\label{eq8}
\textcolor{black}{\mathbf{J}^{\mu}}(\mathbf{r_{1}})=c\textcolor{black}{\overline{\psi}_{f}(\mathbf{r_{1}})\gamma^{\mu}\psi_{i}(\mathbf{r_{1}})},
\end{equation}
where $\psi_{f} $ is the semi relativistic Coulomb wave function~\cite{a35},
\begin{equation}\label{eq9}
\psi_{f}(\mathbf{r_{1}})=N_1\phi_{k1}(Z,\mathbf{r_{1}})\textcolor{black}{u_{\textbf{k}_{1},\mu_{1}}},
\end{equation} 
where $N_1=[1+\frac{{k_1}^2}{4c^2}]^{-1/2} $ and $\phi_{k_{1}}(Z,\mathbf{r_{1}}) $ is Coulomb wave-function defined as,
\begin{eqnarray} \label{eq10} \nonumber 
\phi_{k_{1}}(\textcolor{black}{Z},\textcolor{black}{\textbf{r}_{1}})=\frac{1}{(2\pi)^{3/2}}e^{i\textbf{k}_{1}\cdot\textbf{r}_{1}}\exp\Big(\frac{\pi Z}{2k_{1}}\Big)\Gamma\Big(1+i\frac{Z}{k_{1}}\Big)\\
 {_1F_1}\Big(\frac{-iZ}{k_{1}},1,-i(k_{1}r_{1}+ \textbf{k}_{1}\cdot\textbf{r}_{1})\Big),
\end{eqnarray}
where $Z$ is atomic number.

While, $\psi_{i}(r_{1}) $ is \textcolor{black}{the} Darwin wave function for \textcolor{black}{the} K-shell electrons~\cite{a37}\textcolor{black}{.} \textcolor{black}{It is of the form};
\begin{equation}\label{eq11}
\psi_{i}(r_{1})=N_k\textcolor{black}{\frac{Z^{\prime^{3/2}}}{\sqrt{\pi}}}[\textcolor{black}{a_{s_{b}}(\mu_{b})}]e^{-Z^{\prime} r_{1}}.
\end{equation}
Here, $N_k=[1+\frac{\textcolor{black}{Z^{\prime}}{\alpha}^2}{4}]^{-1/2} $, $\alpha=\frac{1}{c}=\frac{1}{137}$ (in \textcolor{black}{atomic} unit) being the fine structure constant and $Z^{\prime}=Z-0.3 $ is the effective nuclear charge. The Darwin matrix~\cite{a37} \textcolor{black}{$a_{s_{b}} $ is of} the following form;
\begin{equation}\label{eq12}
a_{s_{b}}(\uparrow)=
\begin{pmatrix}
1\\
0\\
\frac{1}{2\textit{i}c}\frac{\delta}{\delta z}\\
\frac{1}{2\textit{i}c} (\frac{\delta}{\delta x}+\textit{i}\frac{\delta}{\delta y})
\end{pmatrix},
\end{equation}

\begin{equation}\label{eq13}
a_{s_{b}}(\downarrow)=
\begin{pmatrix}
0\\
1\\
\frac{1}{2\textit{i}c} (\frac{\delta}{\delta x}-\textit{i}\frac{\delta}{\delta y})\\
\frac{1}{2\textit{i}c}\frac{\delta}{\delta z}
\end{pmatrix}.
\end{equation}
The advantage of the \textcolor{black}{present} FBA model is that it makes \textcolor{black}{our} approach analytic \textcolor{black}{for the computation of} the $\langle f|\widehat{S}|i\rangle$ matrix element. In literature, the Darwin wave function was used by numerous group in the description of the inner shell ionization process on atoms (e.g, see Davidovic and Moiseiwitch(1975)~\cite{a43}, Sud and Moattar(1990)~\cite{a48}, Jacubassa-Amundsen(1989)~\cite{a44}, Bhullar and Sud(2001)~\cite{a35} etc.). Jacubassa-Amundsen(1992)~\cite{a45} used an improved hydrogenic relativistic wave function by replacing the Darwin wave function. It has been observed that it improves the results of (e,2e) processes considerably for Au target. In the present communication, we use Cu and Ag targets wherein, as per the conclusion made by Jacubassa-Amundsen(1992), the improved bound state wave function doesn't lead to a significant change to the Darwin wave function based results. The matrix element for charge-charge interaction can be described as;
\begin{equation}\label{eq14}
\langle f|\widehat{S}|i\rangle_{0} =-\frac{1}{c}\int A_{0}(r_{1})J^{0}(r_{1})d^{3}r_{1}. 
\end{equation}
Similarly, the matrix element for current-current interaction becomes
\begin{equation}\label{eq15}
\langle f|\widehat{S}|i\rangle_{J}=-\frac{1}{c}\int \mathbf{A}(\mathbf{r_{1}})\cdot\mathbf{J}(\mathbf{r_{1}})d^{3}r_{1}. 
\end{equation}
 The total contribution $\langle f|\widehat{S}|i\rangle $ can be written as 
\begin{equation}\label{eq16}
\langle f|\widehat{S}|i\rangle=\langle f|\widehat{S}|i\rangle_{0}-\langle f|\widehat{S}|i\rangle_{J}.
\end{equation}
We can compute $(TDCS)_{0}$, $(TDCS)_{J}$  and $(TDCS)_{T}$ respectively from Eq.~\eqref{eq14}, \eqref{eq15} and \eqref{eq16} by squaring the corresponding S-matrix element.

In the calculation of matrix element $\langle f|\widehat{S}|i\rangle $, we encounter certain types of spatial integrals, as listed in the article~\cite{a34}\textcolor{black}{,} which \textcolor{black}{are} analytical \textcolor{black}{in nature}. In the calculation of TDCS, we consider all 16 possible combinations of spins of participating electrons. We separately calculate TDCS with right-handed \textcolor{black}{helicity} $[TDCS(\lambda=+\frac{1}{2})]$ and left-handed \textcolor{black}{helicity} $[TDCS(\lambda=-\frac{1}{2})] $ of the incident electron. The unpolarized TDCS can be calculated as;
\begin{equation}\label{eq17}
(TDCS)_{unpolarized}=\frac{1}{2}[TDCS(+\frac{1}{2})+TDCS(-\frac{1}{2})]
\end{equation}

\subsection{Twisted electron ionization}\label{sec2b}
After having briefly discussed the basic theory used to describe the relativistic (e,2e) processes for a plane-wave electron beam, we describe the same for a twisted electron beam. Starting with Bessel beams\textcolor{black}{,} we explain the twisted electron wavefunction and calculation of scattering amplitude. Twisted electron beams are conventionally different from plane-wave \textcolor{black}{beam}. They consist of a superposition of plane waves (such as a Bessel beam) with a defined projection of the orbital angular momentum onto the propagation axis\textcolor{black}{,} each with a $\phi$- dependent phase. This phase leads to the characteristic twisted beam with OAM defined by the operator $\hat{L}_{z}=-i\hbar\partial_{\phi} $ with eigenvalue $\hbar \, l $ 
\textcolor{black}{, $l$ is the OAM projection}~\cite{a38}. For the non-relativistic case\textcolor{black}{,} since the Hamiltonian $H$ commutes with the $\hat{L}_{z}$ , $l$ is a good  quantum number and the twisted electron wave-function is an eigenstate of both $H$ and $\hat{L}_{z}$ . However, for the relativistic energy, the Dirac Hamiltonian for the free electron commutes only with the total angular momentum operator $\hat{J}_{z}=\hat{L}_{z}+\hat{\Sigma}_{z}$~\cite{a36}, where $\hat{\Sigma}_{z}$ is the spin operator with eigenvalue $\hbar \, s$. Therefore, in our case\textcolor{black}{,} TAM is a good quantum number, defined as an eigenvalue of the operator $\hat{J}_{z}$ with eigenstate $\psi_{k,m}$, i.e., $\hat{J}_{z}\psi_{k,m}=m\hbar\,\psi_{k,m}$. The simplest form of Bessel beam is provided by the solution of the Schr\"odinger equation in cylindrical coordinates~\cite{a28,a39}. For the non relativistic case\textcolor{black}{,} this exact solution enclose\textcolor{black}{s} the beam features which are: the quantized (projected) OAM $\hbar\, l $ and the longitudinal and transverse momenta $ \hbar\, k_{z}$ and $\hbar \, k_{\perp} $. The form of Bessel beam is\textcolor{black}{;} 
\begin{equation}\label{eq19}
\psi_{\mathbf{k_i},\textcolor{black}{l}}(\textbf{r})=\frac{\sqrt{\varkappa}e^{i\textcolor{black}{l}\phi_r}}{\sqrt{2\pi}}J_{\textcolor{black}{l}}(\textcolor{black}{\varkappa}r_{\perp}){e^{ik_{z}z}}.
\end{equation}
where $\phi_r$ is the azimuthal angle of $\textbf{r}$, $\mathbf{k_{\perp}}$ is the transverse momentum, $ \textcolor{black}{\varkappa= \vert\mathbf{k_{\perp}}\vert}$ is the absolute value of the transverse momentum. In terms of its momentum components, the Bessel beam shows that this state is a ring of tilted plane waves in momentum space. This representation has been used to calculate the elastic Coulomb scattering amplitude~\cite{a39} and further in the potential scattering of electrons in a framework of the generalized Born approximation~\cite{a28}. For the twisted electron part, we use the same formalism as used in the last section for plane wave ionization except that we replace the plane wave for the incident electron with a twisted electron beam. We describe the momentum vector, $\mathbf{k_i}$, of the incident electron as;
\begin{equation}\label{eq20}
\mathbf{k_i} = (k_i\sin\theta_p\cos\phi_p)\hat{x} + (k_i\sin\theta_p\sin\phi_p)\hat{y} + (k_i\cos\theta_p)\hat{z},
\end{equation}
where $\theta_p$  and $\phi_p$ \textcolor{black}{are} the polar and azimuthal angles of $\mathbf{k_i}$ assuming that the electron beam propagates along the \textit{z}-direction.

\begin{figure}[ht]
\includegraphics[width = 0.9\columnwidth]{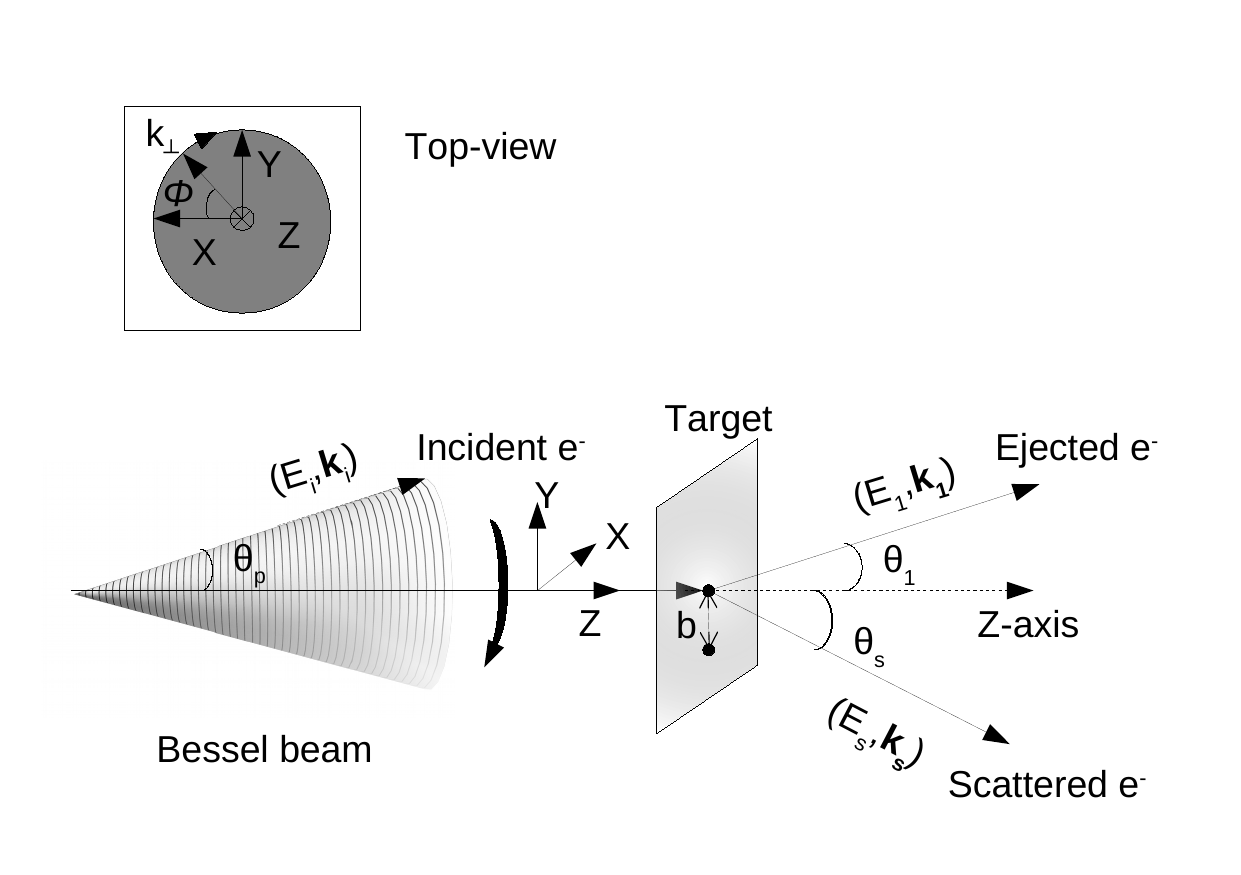}
\caption{Schematic diagram for the electron impact ionization of an atom by Bessel beam in which the momentum of the beam is on the surface of a cone with opening angle $\theta_p$. The angular positions of the scattered and ejected electron are represented by $\theta_s$ and $\theta_1$. The quantization (\textit{z})-axis is chosen along the propagation direction of the incoming beam. Inset shows the top view of the incident twisted electron beam. The beam propagates into the page and twists around the propagation direction (clockwise) with phase angle $\phi_p$. Other kinematical conditions are same as shown in the Fig.~\ref{fig1}.}\label{fig2}
\end{figure}
  
Here, we define the longitudinal momentum along the \textit{z}-axis, with the atom\textcolor{black}{ic target} on the axis. Note that the polar angle is also defined as the opening angle $\theta_p = \tan^{-1}\frac{k_{i\perp}}{k_{iz}}$, which can be defined as the angle that the momentum vector makes with the \textit{z}-axis. $k_{i\perp}$ and $k_{iz}$ represent the perpendicular and the longitudinal components of the momentum  $\mathbf{k_i}$.

We use the Bessel beam in terms of its momentum components. \textcolor{black}{It is} defined as a superposition of the plane waves~\cite{a25},
\begin{equation}\label{eq21}
\psi_{\mathbf{k_{i}},m}(\mathbf{r_i})  = \int \frac{d^2\mathbf{k_{i\perp}}}{(2\pi)^2} a_{\varkappa m}(k_{i\perp})e^{i\mathbf{k_i}\cdot\mathbf{r_i}},
\end{equation}
with the amplitude 
\begin{equation}\label{eq22}
a_{\varkappa m}(k_{i\perp}) = (-i)^m \sqrt{\frac{2\pi}{\varkappa}} e^{im\phi_p} \ \delta(\textcolor{black}{k_{i\perp}} - \varkappa),
\end{equation}
where $\varkappa$ is the absolute value of the transverse momentum, $m$ is TAM projection. For the relativistic case, the wave function for the twisted electron can be described as a superposition of the Dirac plane wave. The wave function for the twisted electron beam can be defined as: 
\begin{equation}\label{eq23} 
\textcolor{black}{\psi^{(tw)}_{\varkappa m}(\mathbf{r_i})  = \int \frac{d^2\mathbf{k_{i\perp}}}{(2\pi)^2} a_{\varkappa m}(k_{i\perp})u_{\mathbf{k_{i}},\lambda}e^{i\mathbf{k_i}\cdot\mathbf{r_i}}},
\end{equation}
Using the Eq.(\ref{eq21}) and (\ref{eq23}) for \textcolor{black}{the} incident twisted electron, we can write the twisted wave transition amplitude, $S^{tw}_{fi}$, in terms of the plane wave transition amplitude as, 
\begin{equation}\label{eq24}
S^{tw}_{fi}(\varkappa,\mathbf{q}) = \frac{(-i)^m}{2\pi}\sqrt{\frac{\varkappa}{2\pi}}\int \frac{d\phi_p}{2\pi}\ e^{im\phi_p - i\textbf{k}_{i\perp}\cdot\textbf{b}}\langle f|\widehat{S}|i\rangle.
\end{equation}
where $\textbf{b}$  is the impact parameter that defines the transverse orientation of the incident twisted electron beam with respect to the atom and $\textbf{k}_{i\perp}\cdot\textbf{b} = \varkappa b \cos(\phi_p - \phi_b)$ and $\phi_b$ is the azimuthal angle of the impact parameter $\mathbf{b}$. $\langle f|\widehat{S}|i\rangle$ is the transition amplitude for the incident plane wave scattering for an atom, given by Eq.~\eqref{eq3}, for the given $\textbf{q}=\textbf{k}_{i}-\textbf{k}_{s} $.

In terms of magnitude, the momentum transfer to the target atom by an incident twisted electron beam is  
\begin{equation}\label{eq25}
q^2 = k_i^2 + k_s^2 - 2k_ik_s\cos\theta,
\end{equation}
where 
\begin{equation}\label{eq26}
\cos\theta = \cos\theta_p\cos\theta_s + \sin\theta_p\sin\theta_s\cos(\phi_p - \phi_s).
\end{equation}
Here, $\theta_s$ and $\phi_s$ are the polar and azimuthal angles of the $\mathbf{k_s}$. For the coplanar geometry, $\phi_s = 180^{\circ}$. For the computation of the TDCS for the twisted electron, we need to compute $\langle f|\widehat{S}|i\rangle$ from Eq.~\eqref{eq3} for given $\phi_p$. We use Eq.~\eqref{eq24} to compute $S^{tw}_{fi}(\varkappa,\mathbf{q})$ by integrating over angle $\phi_p$.
\\ 

Till now\textcolor{black}{,} we have discussed the \textcolor{black}{theoretical formalism for the} (e,2e) cross section for the impinging twisted electron colliding with a single and well localized atomic target. However, \textcolor{black}{this type of} experimental realization \textcolor{black}{is hard to achieve.} \textcolor{black}{In the experiment,} thin foils are usually used\textcolor{black}{.} \textcolor{black}{Such type of} solid target\textcolor{black}{s} can be described by an ensemble of randomly and uniformly distributed identical atoms over the transverse extent of the incident beam. In this case, we can find the average TDCS ($(TDCS)_{av}$), $\frac{\overline{d^3\sigma}}{d\Omega_s\, d\Omega_1 dE_1}$, after integrating the number of events over all the impact parameter $\mathbf{b}$ and dividing it by the total number of particles (Karlovet et al.~\cite{a26}). The average differential cross section can be found as,
\begin{eqnarray}\label{eq27}\nonumber
\frac{\overline{d^3\sigma(\lambda)}}{d\Omega_s \, d\Omega_1 dE_1} = (2\pi)^4 \, \frac{k_S k_1}{k_i}  \,\, \frac{E_i E_S E_1}{c^6} \times \\
\overline{\sum_{\textcolor{black}{\mu_b}}} \sum_{\textcolor{black}{\lambda_s, \mu_1}} \int_{0}^{2\pi} |\langle f|\hat{S}|i\rangle|^2\, \frac{d\phi_p}{2\pi\cos\theta_p}.
\end{eqnarray}

where $\langle f|\hat{S}|i\rangle$ stands for the plane wave S-matrix element as defined by Eq.~\eqref{eq16}. Note that the average TDCS does not depend on the TAM \textcolor{black}{projection} $m$. However, it depends on the opening angle $\theta_p$ of the incident twisted electron. 

We also calculate the asymmetry $A_{L}$ in $(TDCS)_{av}$ as;
\begin{equation}\label{eq18}
A_{L}= \frac{(TDCS)_{av}(+\frac{1}{2}) - (TDCS)_{av}(-\frac{1}{2})}{(TDCS)_{av}(+\frac{1}{2})+(TDCS)_{av}(-\frac{1}{2})}.
\end{equation}
The spin asymmetry in \textcolor{black}{the} K-shell ionization \textcolor{black}{process} is caused by the spin-dependent forces, i.e., by Mott scattering (due to the spin orbit interaction of the continuum electrons moving with relativistic energies in the Coulomb field of the atomic nucleus).

\textcolor{black}{F}or the non-relativistic energy case\textcolor{black}{, we} benchmark our calculation with \textcolor{black}{that of} Harris et al. (2019) work. For the twisted electron \textcolor{black}{beam imapct (e,2e) process}, the matrix element $T_{fi}^{tw}$ can be expressed as~\cite{a31},
\begin{equation}\label{eq28}
T_{fi}^{tw} = \frac{(-1)^l}{2\pi} \int_{0}^{2\pi} d\phi_p \,\, e^{i l \phi} \, T_{fi}^{PW}(\mathbf{q}) \,\, e^{-i \textcolor{black}{\mathbf{k_i}}\cdot \mathbf{b}}.
\end{equation}
Here, $l$ is the orbital angular momentum (OAM) \textcolor{black}{projection} \textcolor{black}{of the twisted electron beam}. Further, the average TDCS can be calculated by integrating the $|T_{fi}^{tw}(\mathbf{q})|^2$ over the impact parameter $\mathbf{b}$ in the transverse plane. 

The average cross section can be expressed as:

\begin{equation}\label{eq29}
\frac{\overline{d^3 \sigma}}{d\Omega_s \, d\Omega_1 dE_s}= \frac{1}{2\pi \,\cos \theta_p} \int_{0}^{2\pi} d\phi_p \Big(\frac{d^3\sigma(\mathbf{q})}{d\Omega_s \, d\Omega_1 dE_s}\Big)_{PW}.
\end{equation}

\section{Results and Discussion}\label{sec3}
\begin{figure*}[htp]
\includegraphics[width = 0.8\columnwidth]{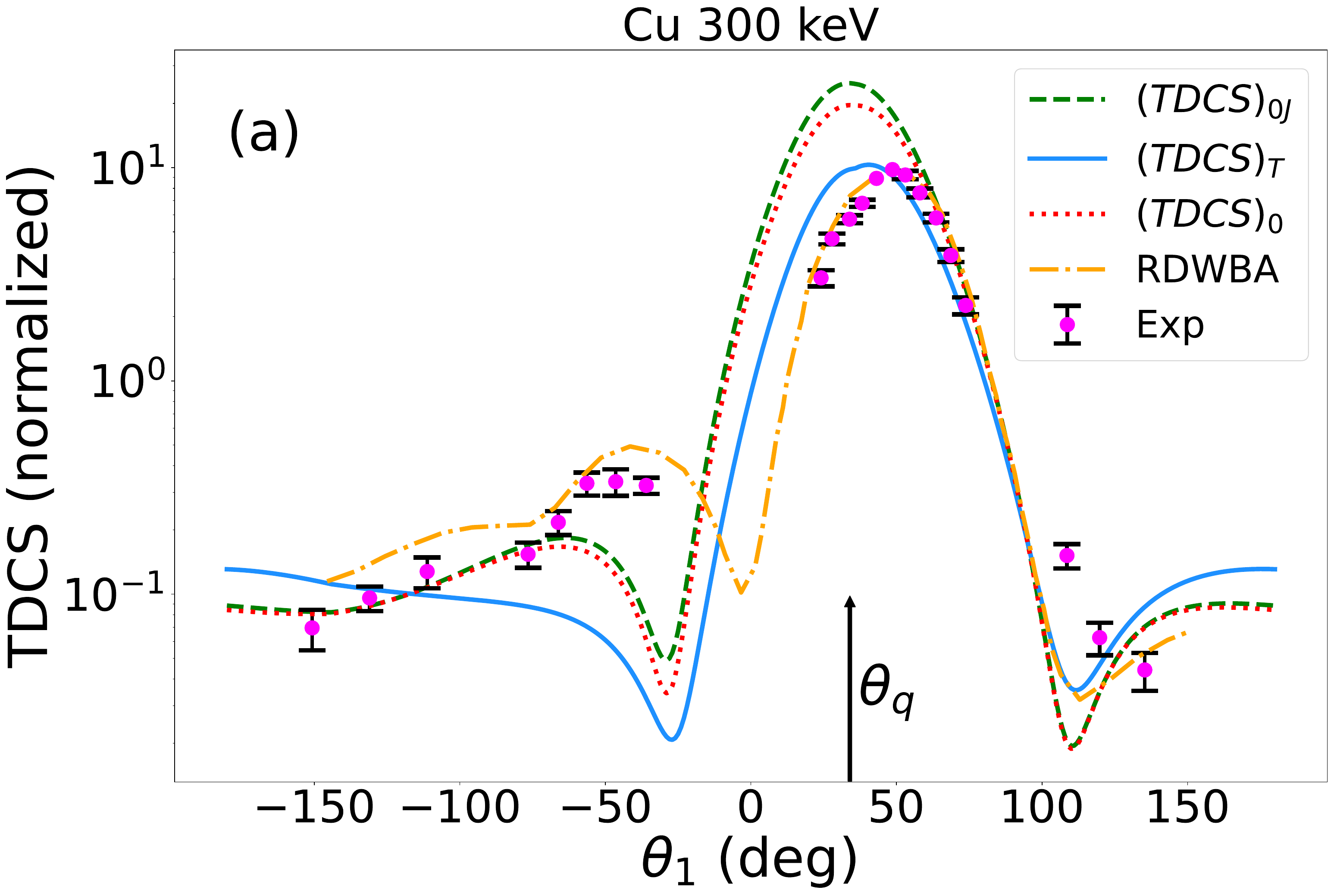}
\includegraphics[width = 0.8\columnwidth]{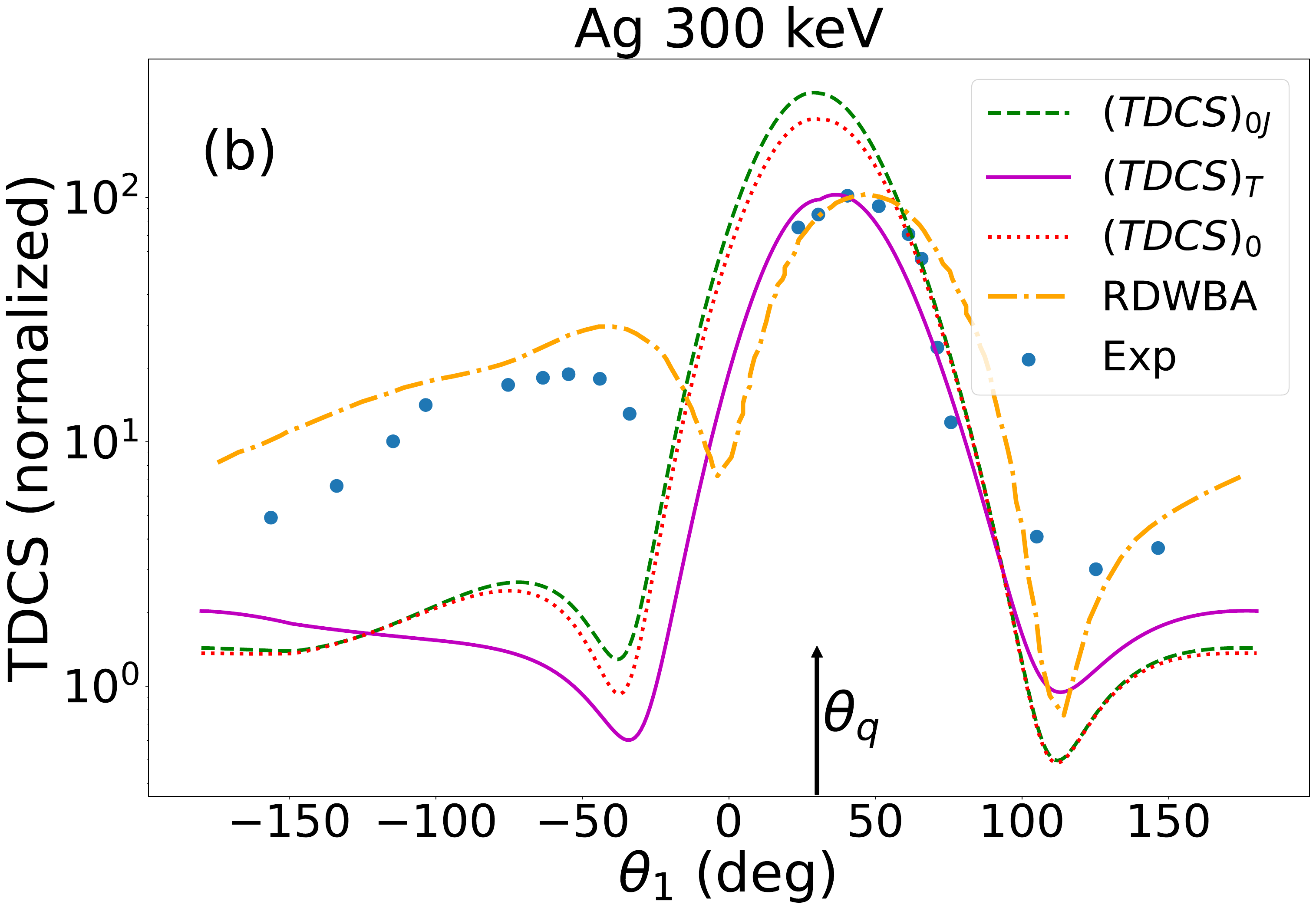}
\includegraphics[width = 0.8\columnwidth]{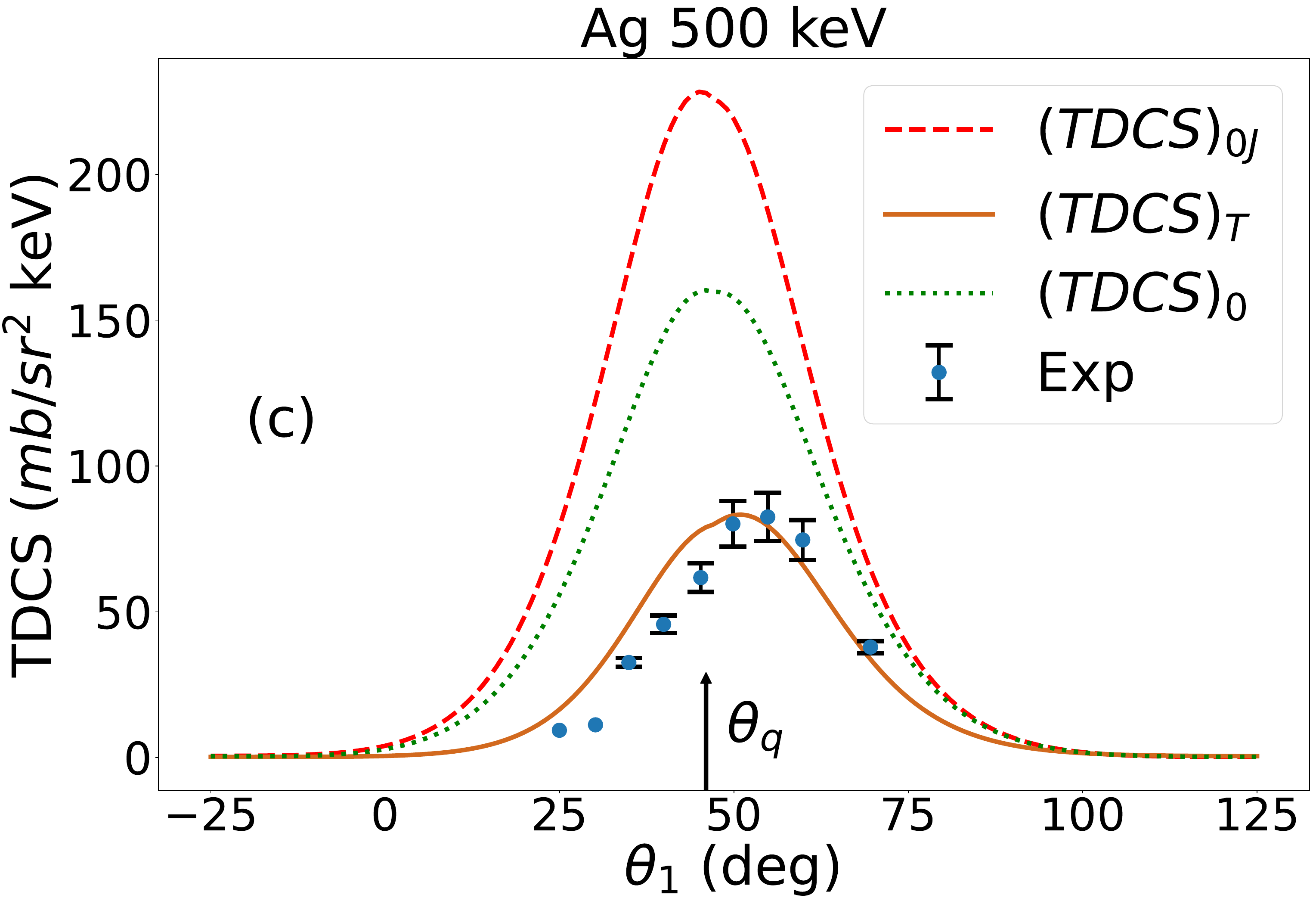}
\caption{$(TDCS)_{T}$, $(TDCS)_{0}$, $(TDCS)_{0J}$, rDWBA calculation are plotted as a function of the ejected electron angle ($\theta_{1}$) in the coplanar asymmetry geometry mode. The experimental data points (from ~\cite{a33},~\cite{a41}) are plotted as a symbol($\bullet$). rDWBA calculation are represented by dot-dashed curve (~\cite{a32,a33}). $(TDCS)_{T}$, $(TDCS)_{0}$, $(TDCS)_{0J}$ are represented by solid,dotted and dashed curves respectively. Kinematics for Fig(a):Cu ($E_{i}=$300 keV, $E_{s}=$220 keV, $E_{1}=$71 keV, $\theta_{s}=9^{\circ}$), Fig(b):Ag ($E_{i}=$300 keV, $E_{s}=$200 keV, $E_{1}=$74.5 keV, $\theta_{s}=10^{\circ}$), Fig(c):Ag ($E_{i}=$500 keV, $E_{s}=$375 keV, $E_{1}=$100 keV, $\theta_{s}=15^{\circ}$). Arrow at $\theta_{q}$ represents the direction of momentum transfer. In all the cases, $\phi_s = 180^{\circ}$.}\label{fig3}
\end{figure*}

We present the results of our calculations of TDCS for incoming plane wave \textcolor{black}{electron beam} in Fig.~\ref{fig3} for the charge-charge interaction $(TDCS)_{0}$, the sum of the charge-charge and current-current interaction part $((TDCS)_{0J}=(TDCS)_{0} + (TDCS)_{J})$ and total contributions\textcolor{black}{,} which includes the interference term of the matrix elements of the above contributions $((TDCS)_{T}=(TDCS)_{0}+(TDCS)_{J}+(TDCS)_{INT})$. We describe $(TDCS)_{0}$, $(TDCS)_{0J}$ \textcolor{black}{and} $(TDCS)_{T}$ by dotted, dashed and solid curves respectively\textcolor{black}{.} \textcolor{black}{The} experimental data \textcolor{black}{are described} by ($\bullet$) and \textcolor{black}{the} rDWBA calculations by dashed-dotted curve. We \textcolor{black}{present the} results \textcolor{black}{of TDCS} for Cu  and Ag targets for \textcolor{black}{an} incident energy $E_{i}=$300 keV (Cu and Ag) and $E_{i}=$500 keV (Ag) in \textcolor{black}{the} coplanar asymmetric geometrical mode. We depict other kinematical variables of the calculation of TDCS in the caption of Fig.~\ref{fig3}. We normalize the experimental and our FBA results to the rDWBA calculation of Keller and Dreizler~\cite{a32} for Cu and that for Keller et al.~\cite{a33} for Ag. We normalize our calculation to rDWBA as it predicts correct cross sections for the present kinematics. We found that for Cu (Fig.~\ref{fig3}(a)), our theoretical results follow the experimental and rDWBA results reasonably well in the binary peak region (see the peak around $\theta_{1}=\theta_{q} $ region, i.e. in the momentum transfer direction). We also note that the angular profile of $(TDCS)_{0}$ (dotted curve) and $(TDCS)_{0J}$ peak around $\theta_{1}=\theta_{q}$ . However, because of the interference of the scattering amplitudes of the charge density and current density terms, the $(TDCS)_{T}$ is reduced and the binary peak is shifted to a higher angle (see solid curve). \textcolor{black}{Further}, we found that the interference term is responsible for the additional small maximum in the backward region (see solid curves in the region near to $\theta_{1}=$ $\pm180^{\circ})$. However, for Ag target, we found that our present FBA calculation differs strongly from the rDWBA and experimental \textcolor{black}{trends} in the recoil peak region (see Fig.~\ref{fig3}(b)). This is expected as our FBA  \textcolor{black}{calculation is} good for the lighter target, like Cu. The FBA \textcolor{black}{calculation is not accurate} for heavier targets. The FBA fails for heavy targets as it neglects the action of \textcolor{black}{the} target field on the projectile electron in both the entrance and exit channels (see Keller et al.(1999)~\cite{a32}). Further, the FBA result in the recoil peak region is poor as it doesn't take into account the scattered wave contribution in the plane wave representation of the incident electron. Despite of this, the FBA calculation in the binary peak region follows the experimental and rDWBA results (see solid and dashed-dotted curves in the binary peak region with their comparison with the experimental data in Fig.~\ref{fig3}(a) and ~\ref{fig3}(b)). In the present scenario, we compare our theoretical results with the experimental data and rDWBA calculation to ascertain that it reproduces the main trends of the angular profile of TDCS of experimental data and rDWBA  model in the dominant binary peak region. \textcolor{black}{In Fig.~\ref{fig3}(c) for Ag target, our theory has been normalised to the absolute TDCS of experimental data by a factor of 8.2 and our results underestimate the plane wave Born approximation (PWBA) result of ~\cite{a41} roughly by a factor of 3.56}.
  
\begin{figure*} [htp]
\includegraphics[width=0.8\columnwidth, clip=true] {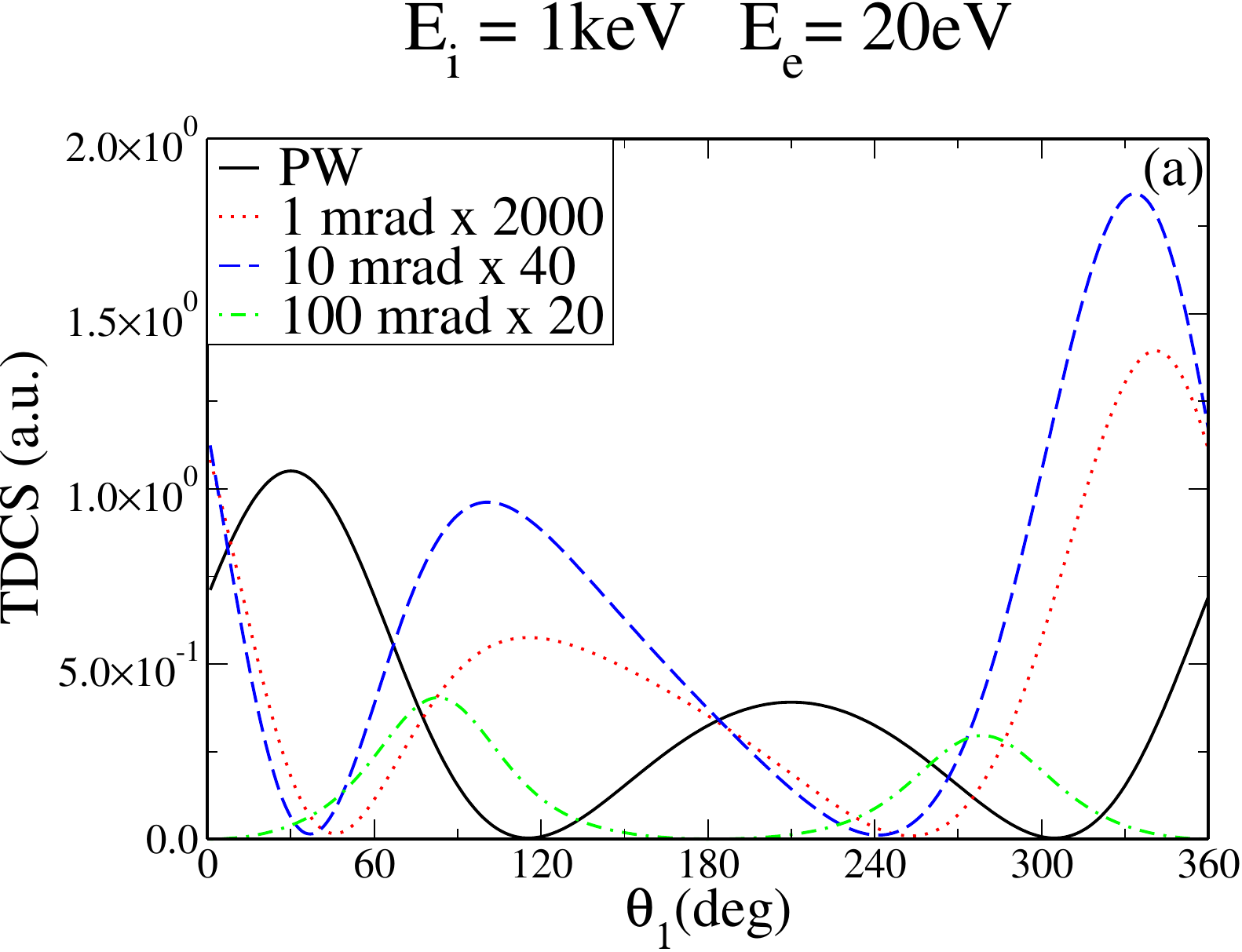} 
\includegraphics[width=0.8\columnwidth, clip=true] {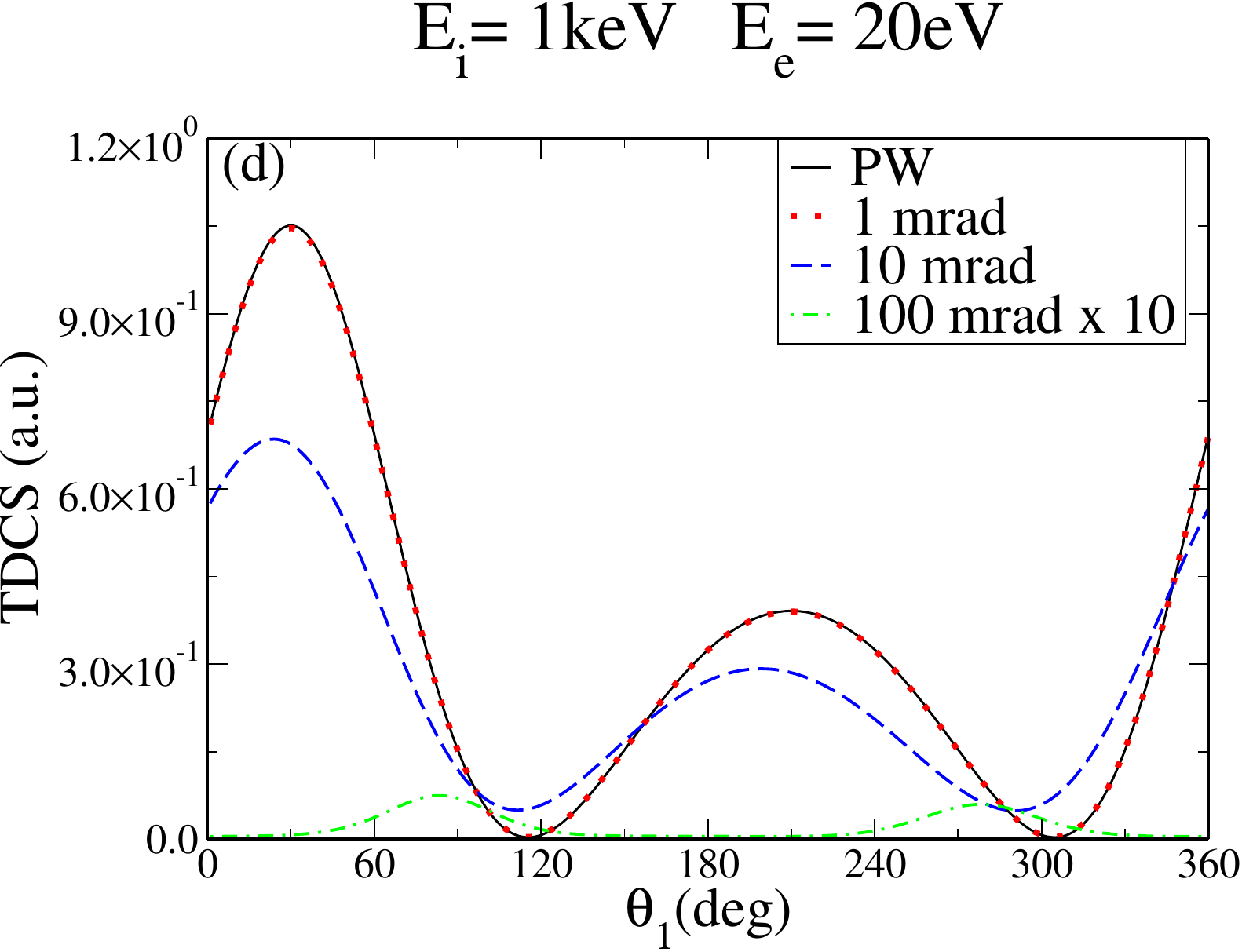} 
\includegraphics[width=0.8\columnwidth, clip=true] {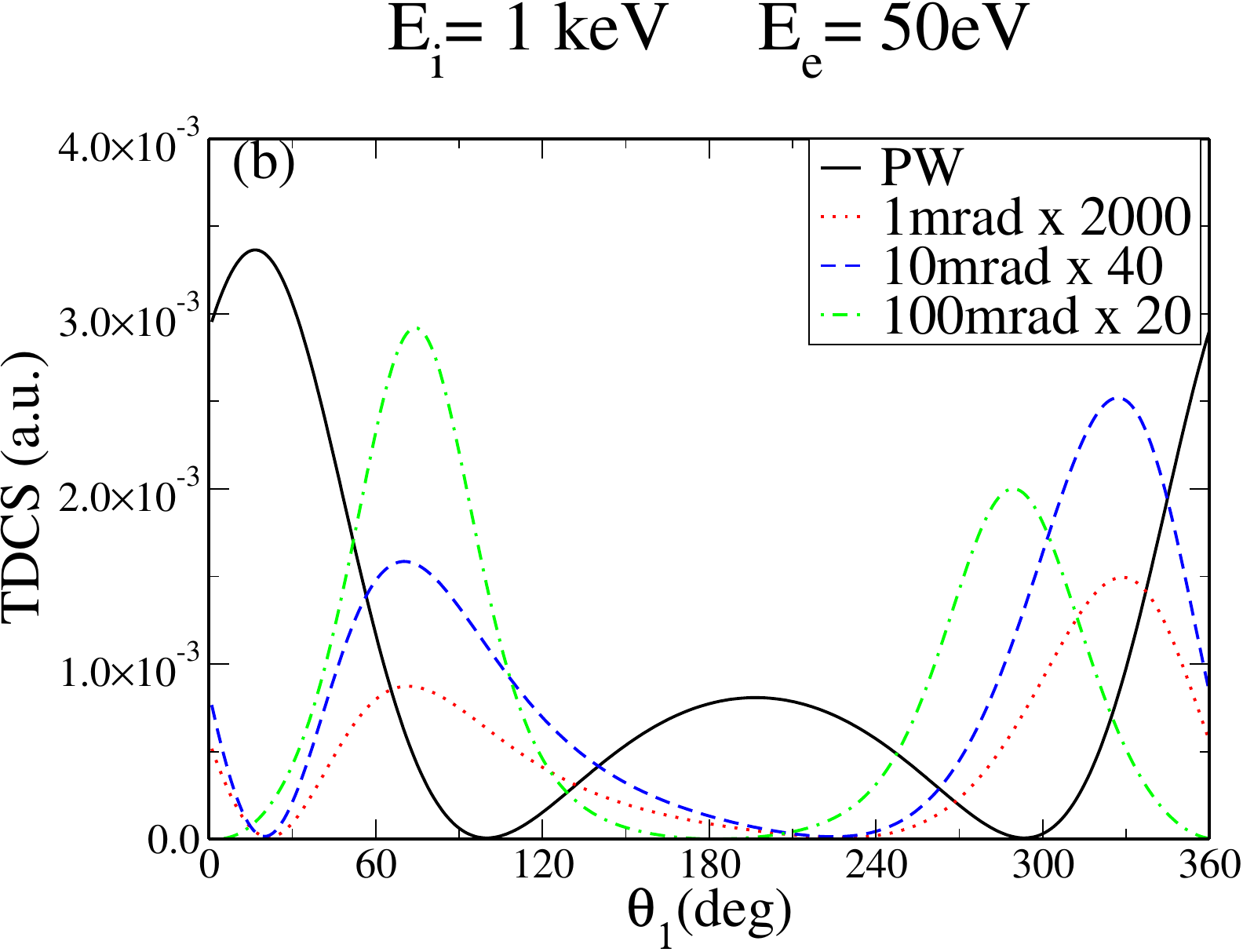}
\includegraphics[width=0.8\columnwidth, clip=true] {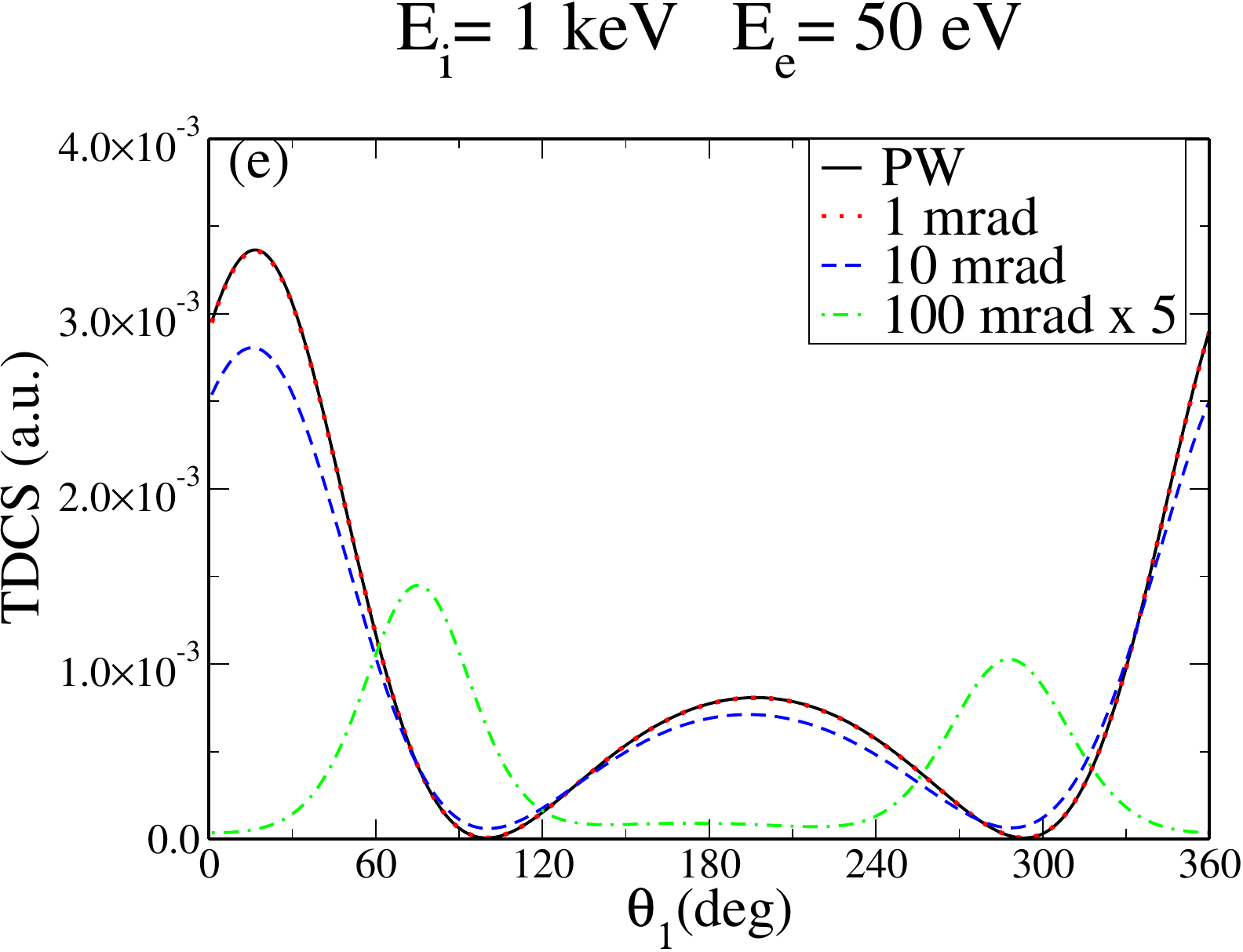}
\includegraphics[width=0.8\columnwidth, clip=true] {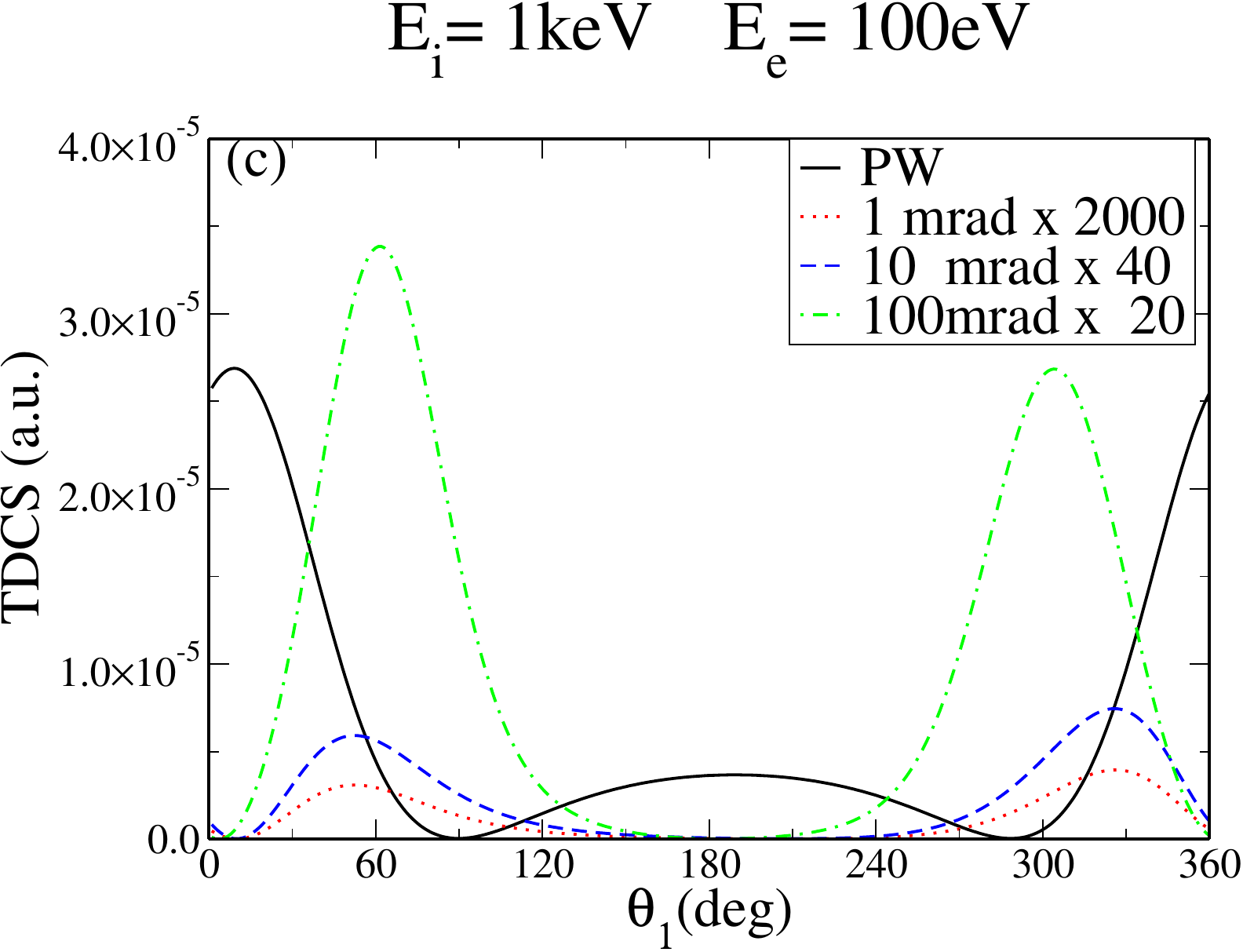}
\includegraphics[width=0.8\columnwidth, clip=true] {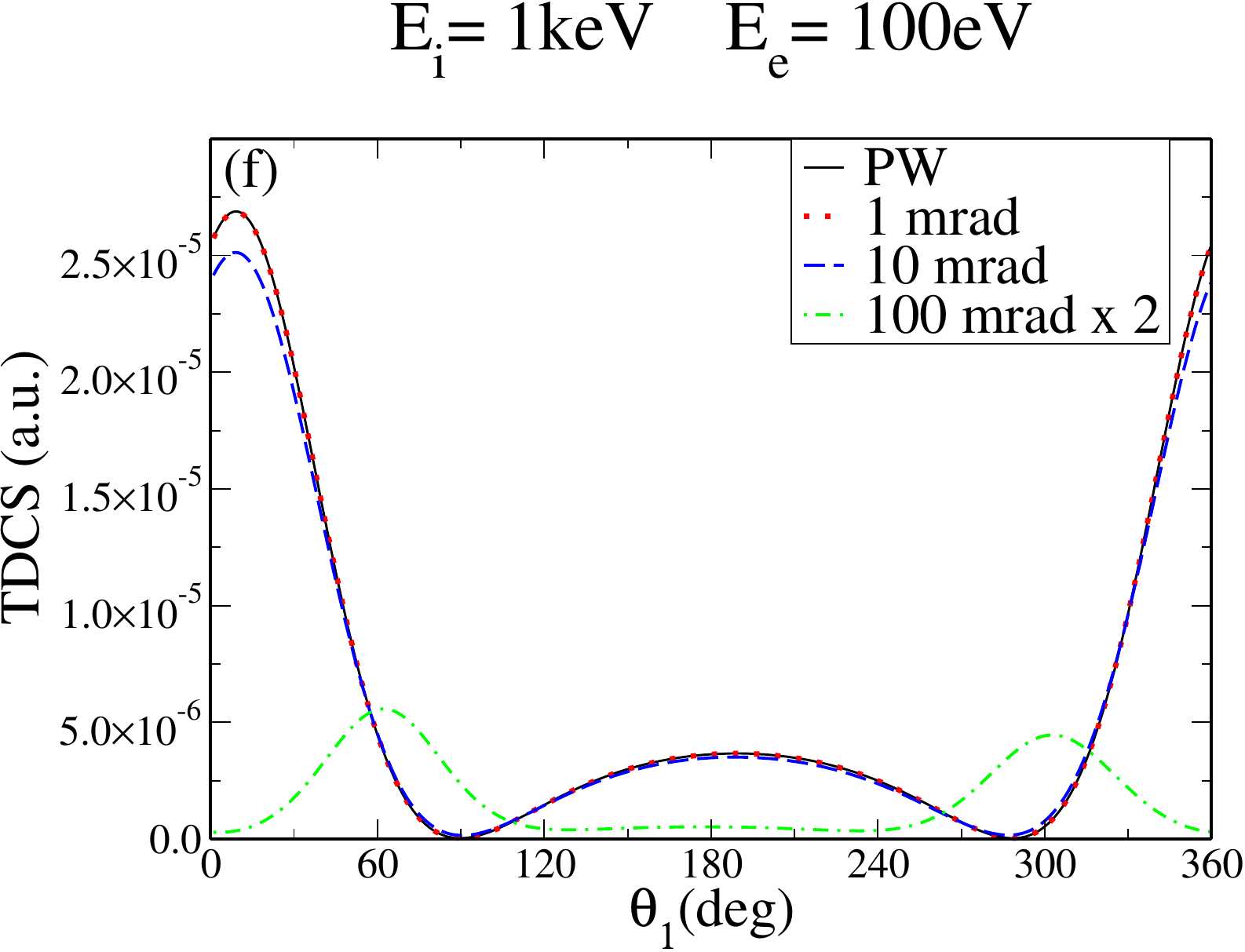}
\caption{TDCS for ionization of hydrogen for the impact energy $E_i=$1 keV plotted as a function of the angle of the ejected electron in the coplanar asymmetric geometry. The solid curve represents plane wave and on-axis electron vortex beam projectiles with opening angles of 1 mrad (dotted line), 10 mrad (dashed line), and 100 mrad (dash dot line) with projectile scattering angle 10 mrad and OAM $l$=1 in frames (a), (b) and (c). Averaged over impact parameter TDCS for ionization of hydrogen for the impact energy $E_i=$1 keV plotted as a function of the angle of the ejected electron in the coplanar asymmetric geometry by plane wave (solid line) and on-axis electron vortex beam projectiles with opening angles of 1 mrad (dotted line), 10 mrad (dashed line) and 100 mrad (dash dot line) with projectile scattering angle 10 mrad and OAM $l$=1 in frames (d), (e) and (f).}\label{fig4}
\end{figure*}

\begin{figure*} [htp]
\includegraphics[width=0.8\columnwidth, clip=true] {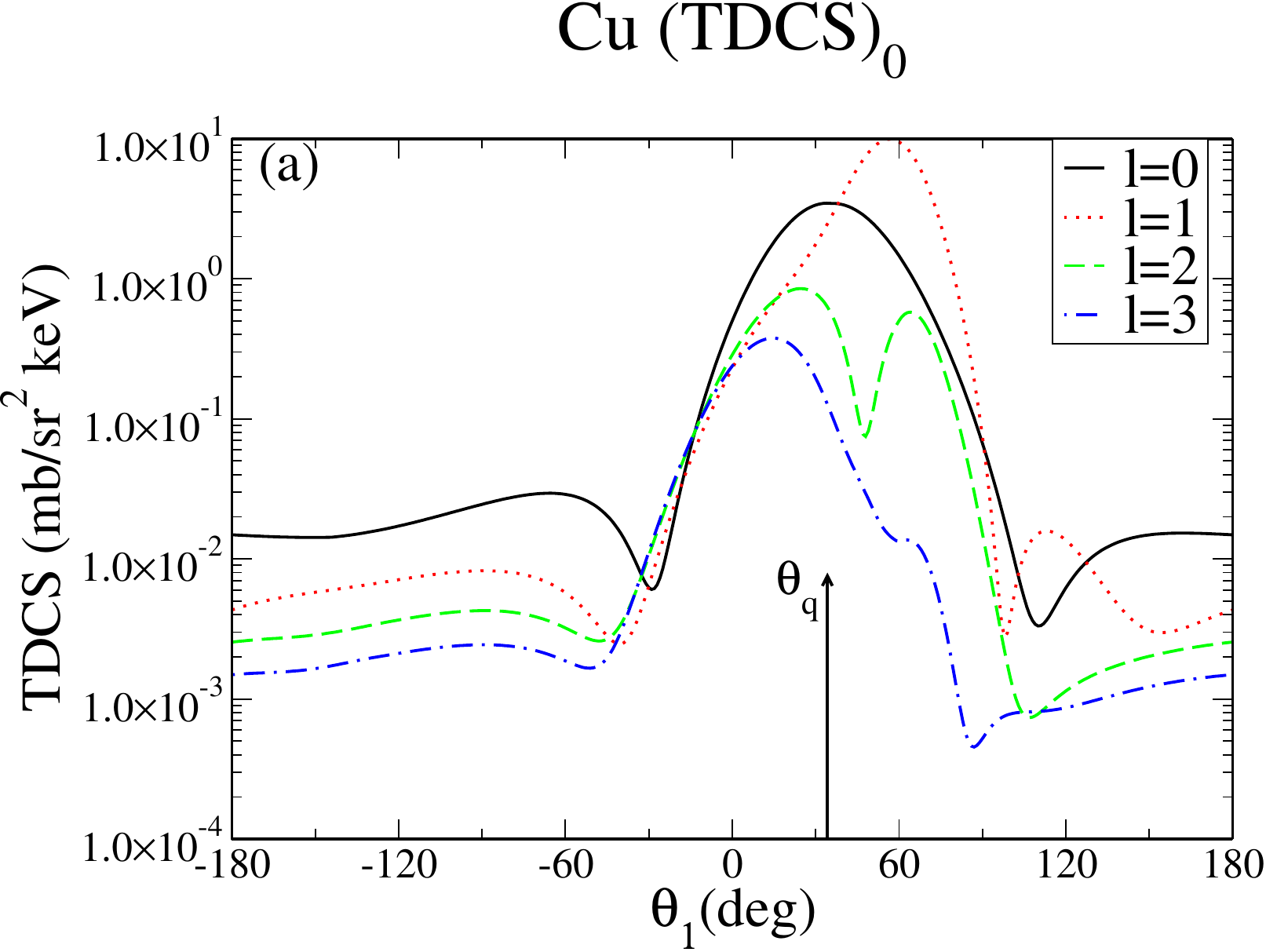}
\includegraphics[width=0.8\columnwidth, clip=true] {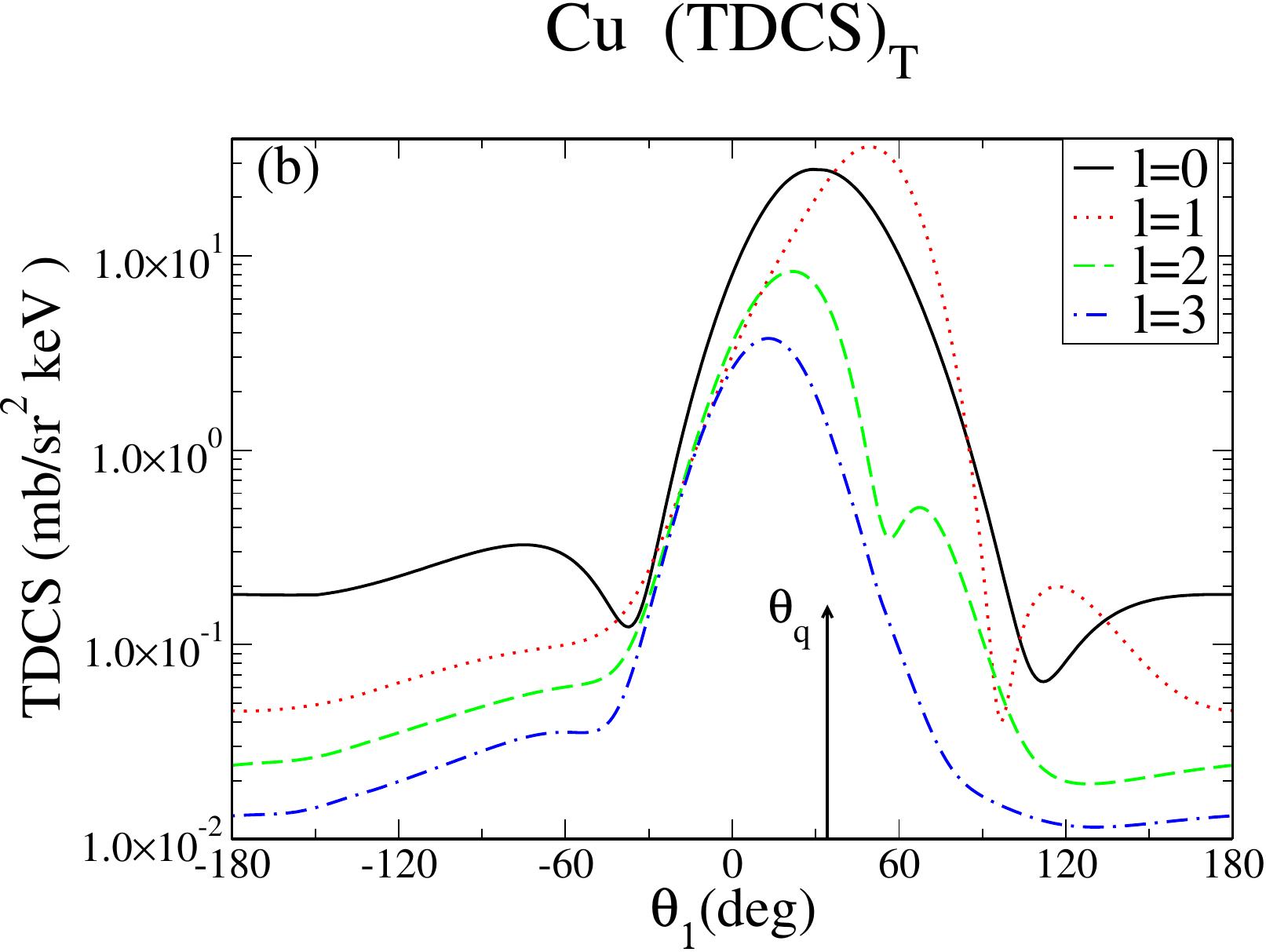}
\includegraphics[width=0.8\columnwidth, clip=true] {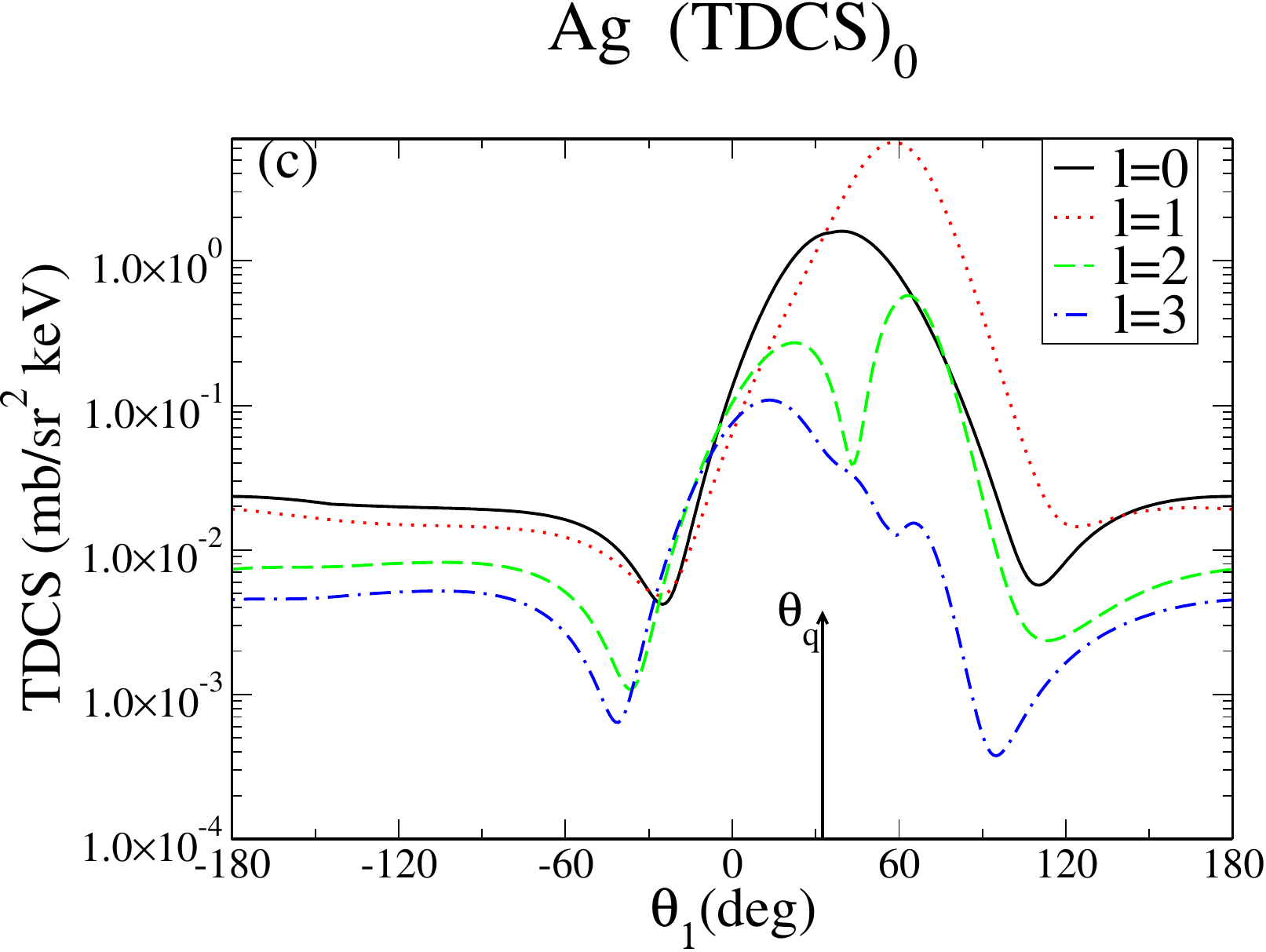}
\includegraphics[width=0.8\columnwidth, clip=true] {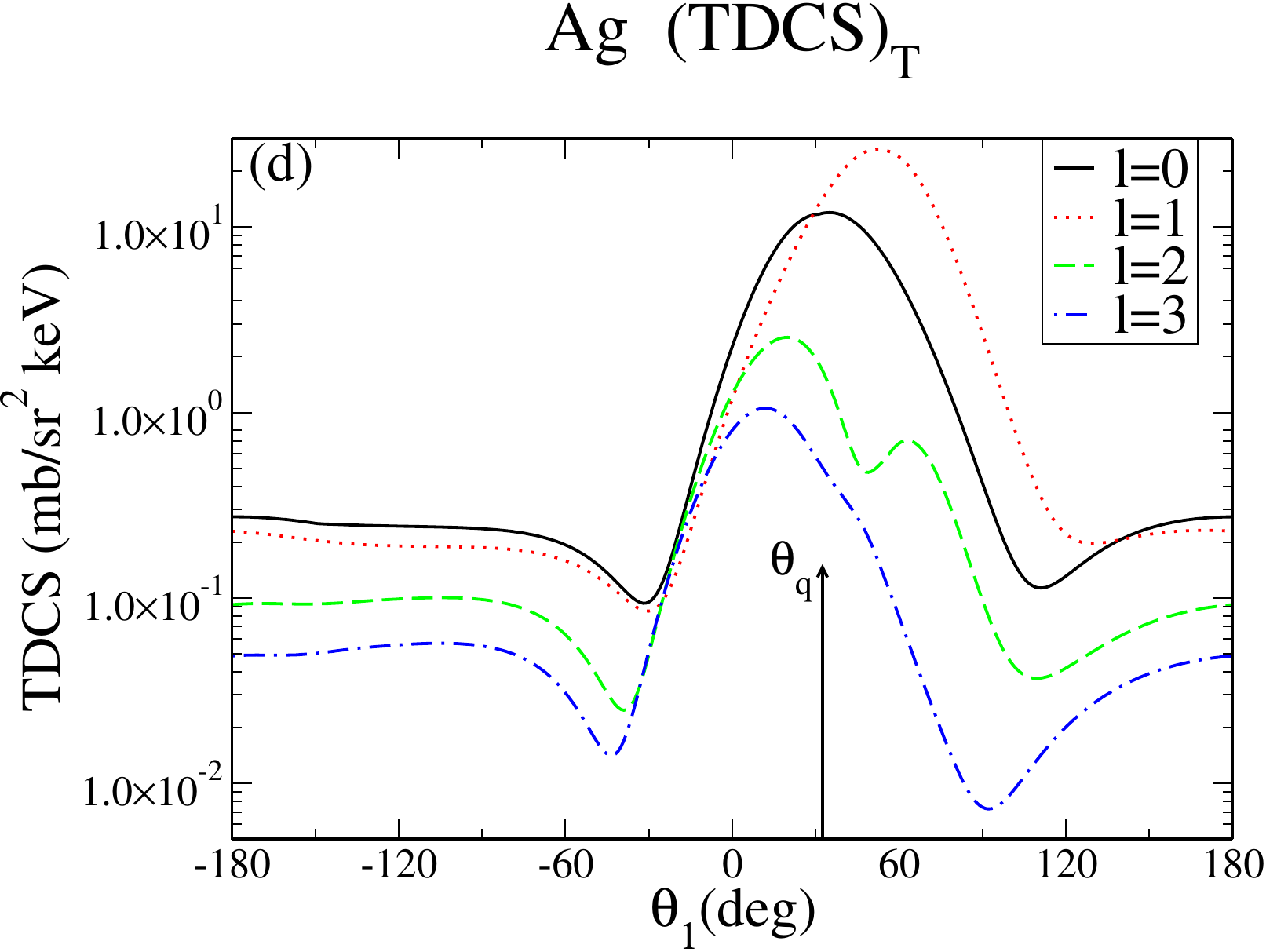}
\caption{$(TDCS)_{0}$ and $(TDCS)_{T}$ for $l$=0, 1, 2 and 3 for Cu($E_{i}=$300 keV, $E_{s}=$220 keV, $E_{1}=$71 keV) with $\theta_{s}=9^{\circ}$ and $\theta_{p}=9^{\circ}$ and Ag($E_{i}=$300 keV, $E_{s}=$200 keV,  $E_{1}=$74.5 keV) with $\theta_{s}=10^{\circ}$ and $\theta_{p}=10^{\circ}$. Fig.5(a) and Fig.5(c) respresent $(TDCS)_{0}$ for Cu and Ag respectively. Similarly, Fig.5(b) and Fig.5(d) respresent $(TDCS)_{T}$ for Cu and Ag respectively.}\label{fig5}
\end{figure*}
 
Before we discuss the effects of TAM \textcolor{black}{projection} $m$ and opening angle $\theta_{p}$ on the relativistic (e,2e) process, we discuss the (e,2e) process on Hydrogen atom at the non-relativistic energy regime. In the literature, Harris et al. (2019)~\cite{a31} discussed the twisted electron impact (e,2e) process on Hydrogen atom for the incident energy 500 eV and 1 keV in the coplanar asymmetric geometry. They took the ejected electron energy as 20 eV, 50 eV and 100 eV with different scattering angles and opening angles of the incident electron (1 mrad, 10 mrad and 100 mrad). Further, they presented the results for two different values of orbital angular momentum (OAM) number, viz $l=1$ and $l=2$. Here, we present the results of our calculation of TDCS for 1 keV impact energy for $\theta_{s} = 10$ mrad with three different values of $\theta_{p}$, i.e. $\theta_{p}= 1$ mrad, 10 mrad and 100 mrad. We choose $l=1$ and the ejected electron's energy as 20 eV, 50 eV and 100 eV. The purpose of the re-calculation of TDCS in the non-relativistic case for Hydrogen is to benchmark our calculation with the existing results of Harris et al. (2019) so that we can justify our \textcolor{black}{semi-relativistic} model  calculation for the relativistic (e,2e) process for the twisted electron beam. \textcolor{black}{Harris et al.(2019) used plane wave for the ejected electron whereas we used Coulomb wave for the ejected electron which is a better wave function for the low energy ejected electron.}

We present the results of our calculation in Fig.~\ref{fig4}. Fig.~\ref{fig4}(a), Fig.~\ref{fig4}(b) and Fig.~\ref{fig4}(c) depict the angular profiles of TDCS of our calculation for the ejected electron energy ($E_{1}$) 20 eV, 50 eV and 100 eV respectively. The scattered electron is detected at a fixed value of 10 mrad. We further compute TDCS for (e,2e) process on Hydrogen by a plane wave and twisted electron by taking the average over impact parameter with opening angles of 1 mrad, 10 mrad, 100 mrad. We keep the scattering angle $\theta_{s}=10$ mrad. The averaged over impact parameter TDCS results mimic the realistic scenario for the present day experimental setup as it is very difficult to do the experiment with a single atom with the twisted electron beam incidence. We depict the results of the TDCS averaged over impact paprameter in Fig.~\ref{fig4}(d), (e) and (f), \textcolor{black}{for $E_{1}$= 20 eV, 50 eV and 100 eV respectively}. 

We confirm that our calculation of TDCS for $\theta_{p}=$1 mrad, 10 mrad, 100 mrad respectively  reproduces the salient features of Harris et al. (2019) paper. The position of the peaks in TDCS for various values of $\theta_{p}$ \textcolor{black}{is} more or less peaked in the directions as predicted by Harris et. al. (2019) (see Fig.~\ref{fig4}(a)-(c)). This is also true for the TDCS averaged over the impact parameter. For example, in the Fig.~\ref{fig4}(d)-(f)\textcolor{black}{,} TDCS for the plane wave (solid curve) almost overlap\textcolor{black}{s} with that \textcolor{black}{of twisted electrons} for $\theta_{p} = 1$ mrad (dotted curve) and almost follow the same pattern as that for $\theta_{p}= 10$ mrad (dashed curve).  We further confirm that the binary and recoil peak for $\theta_{p} = 100$ mrad shifts a lot from their respective position for the plane wave case (see dashed-dotted curve). Our calculated TDCS is smaller than the TDCS of Harris et al. (2019). 

Now, in order to investigate the effects of the different $m$ on the angular profile of spin averaged TDCS in the relativistic (e,2e) process, we present the results of $(TDCS)_{0}$ and $(TDCS)_{T}$ for $l$=0, 1, 2 and 3 in the Fig.~\ref{fig5} for which $m$ can be $|l-1/2|$ and $|l+1/2|$ for given $l$. The spin averaged TDCS can be written as $TDCS=[TDCS(l+1/2)+TDCS(l-1/2)]/2=[TDCS(m)+TDCS(m-1)]/2$. We choose the same targets as used in Fig.~\ref{fig3} with the same kinematics used there for the plane wave (e,2e) case. For each kinematics, we keep the opening angle ($\theta_{p}$) equal to the scattering angle ($\theta_{s}$) with the atom located on the beam direction ($\mathbf{b}=0$). For example, we used $\theta_{s}=9^{\circ}$ for Cu target at 300 keV for the plane wave case. We keep $\theta_{p}=9^{\circ}$ for the present case. We present the results of spin averaged $(TDCS)_{0}$ in Fig.~\ref{fig5}(a) and Fig.~\ref{fig5}(c) and $(TDCS)_{T}$ in Fig.~\ref{fig5}(b) and Fig.~\ref{fig5}(d) for Cu and Ag targets \textcolor{black}{respectively} at $E_{i}=$300 keV incident energy. We use solid, dotted, dashed and dashed-dotted curve\textcolor{black}{s} respectively for $l$= 0, 1, 2 and 3. \textcolor{black}{From now onwards, we} follow the same representation \textcolor{black}{ of different curves in} the paper unless otherwise stated. We plot $(TDCS)_{0}$ and $(TDCS)_{T}$ for different $l$ in different frames to investigate the effects of $l$ (and hence the TAM \textcolor{black}{projection} $m$) on charge-charge interaction $((TDCS)_{0})$ and that for the charge-charge and current-current interaction terms $((TDCS)_{T})$. At relativistic energy, we expect that the current-current interaction term also dominates. Whereas, in the non-relativistic regime, alone $(TDCS)_{0}$ will be sufficient to consider.
 
For 300 keV impact energy, we observe that the binary peak of $(TDCS)_{0}$ shifts \textcolor{black}{from binary peak direction} for $l\neq0$ even for our first Born approximation results. \textcolor{black}{Further, we observe that as $l$ increases, the position of the prominent peak in the binary region shifts to the forward direction for both the targets (e.g., see the peak position of $l$=3 case). For $l$=2, we observe a two peak structure (see dashed curves in Fig. 5(a) and 5(c)). We also observe that the magnitude of   $(TDCS)_{0}$ decreases with $l$. For $l$ = 3, we observe that the binary peak completely vanishes and a prominent peak in the forward direction (see dashed-dotted curves in Fig.~\ref{fig5}(a) and ~\ref{fig5}(c) around $\theta =0^{\circ}$ direction)}. 
 
Having seen the angular profiles of $(TDCS)_{0}$, we now discuss the angular profile of $(TDCS)_{T}$ for $l$=0, 1, 2 and 3 in Fig.~\ref{fig5}(b) and \ref{fig5}(d) for Cu and Ag targets respectively. When we compare $(TDCS)_{T}$ with the $(TDCS)_{0}$ for different values of $l$, we observe that the binary peak of $(TDCS)_{T}$ shifts to larger angles which are similar to what we observe for the plane wave calculation earlier (see Fig.~\ref{fig3}). Apart from this, we observe that the angular profiles of $(TDCS)_{T}$ follow the same patterns as we get for $(TDCS)_{0}$, like the dominant peaks for $l$=0 and $l$=1 are still found for $(TDCS)_{T}$ (see solid and dotted curves of Fig.~\ref{fig5}(b) and Fig.~\ref{fig5}(d)). \textcolor{black}{We again observe dominant peak in the forward direction for larger $l$ (see dashed and dashed-dotted curves in the forward direction in the Fig. ~\ref{fig5}(b) and ~\ref{fig5}(d)) . For $l$ = 2, we observe that the binary peak splits (see dashed curve in the Fig. ~\ref{fig5}(b) and ~\ref{fig5}(d)). For both the targets, we observe a significant change in the angular profile of $(TDCS)_{0}$ when compared with that of $(TDCS)_{T}$ for $l$ = 2 in the binary region (see dashed curves of Fig.~\ref{fig5}(a) and (c) and compare with Fig.~\ref{fig5}(b) and (d))}. We found that the magnitude of $(TDCS)_{0}$ and $(TDCS)_{T}$ decreases as $l$ increases.

\begin{figure}[htp]
\includegraphics[width = 0.8\columnwidth]{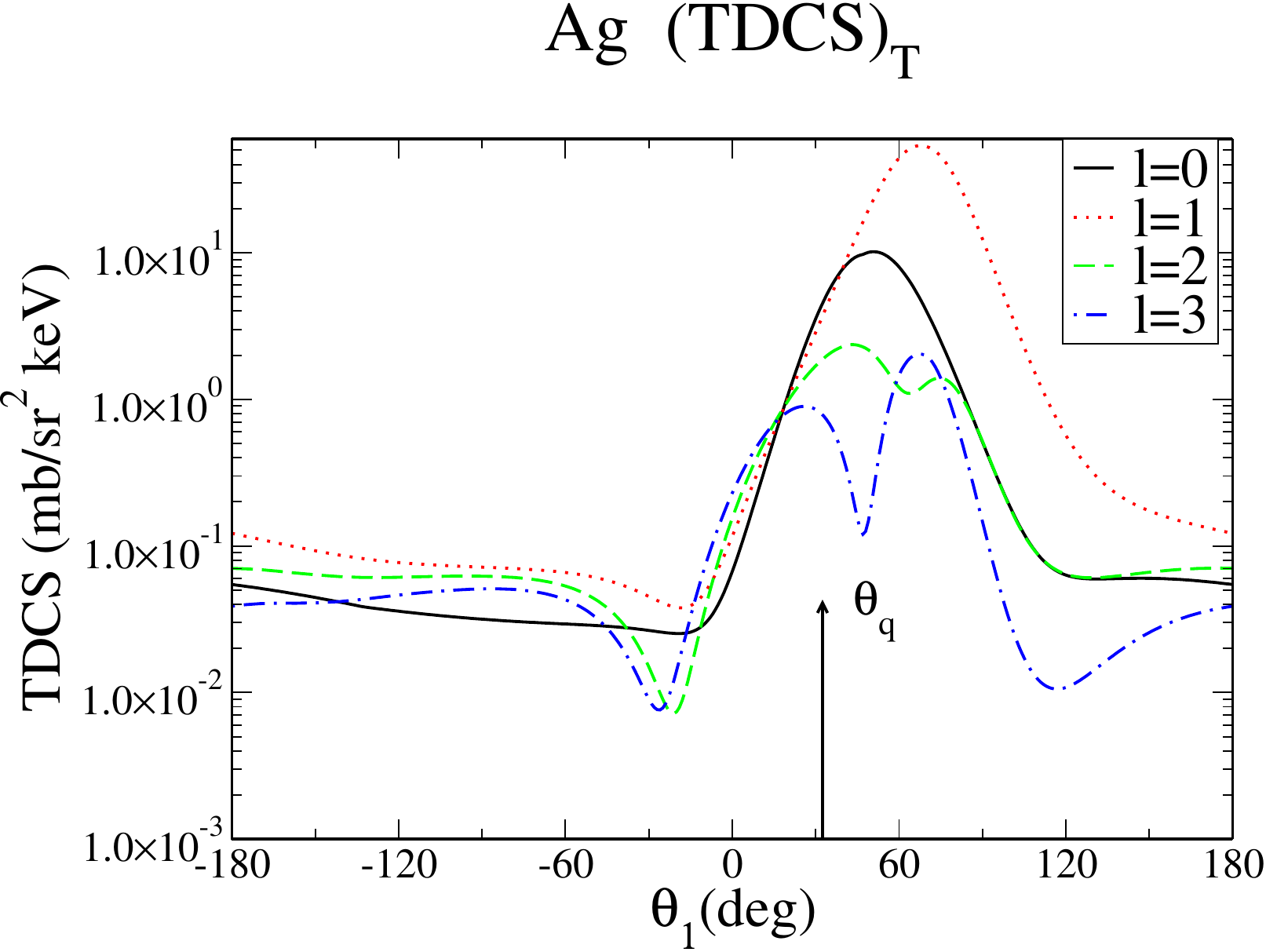}
\caption{$(TDCS)_{T}$ for Ag ($E_{i}=$500 keV,$E_{s}=$375 keV,$E_{1}=$100 keV)for given $l$'s with $\theta_{s}=15^{\circ}$ and $\theta_{p}=15^{\circ}$.}\label{fig6}
\end{figure}

\begin{figure}[htp]
\includegraphics[width = 0.8\columnwidth]{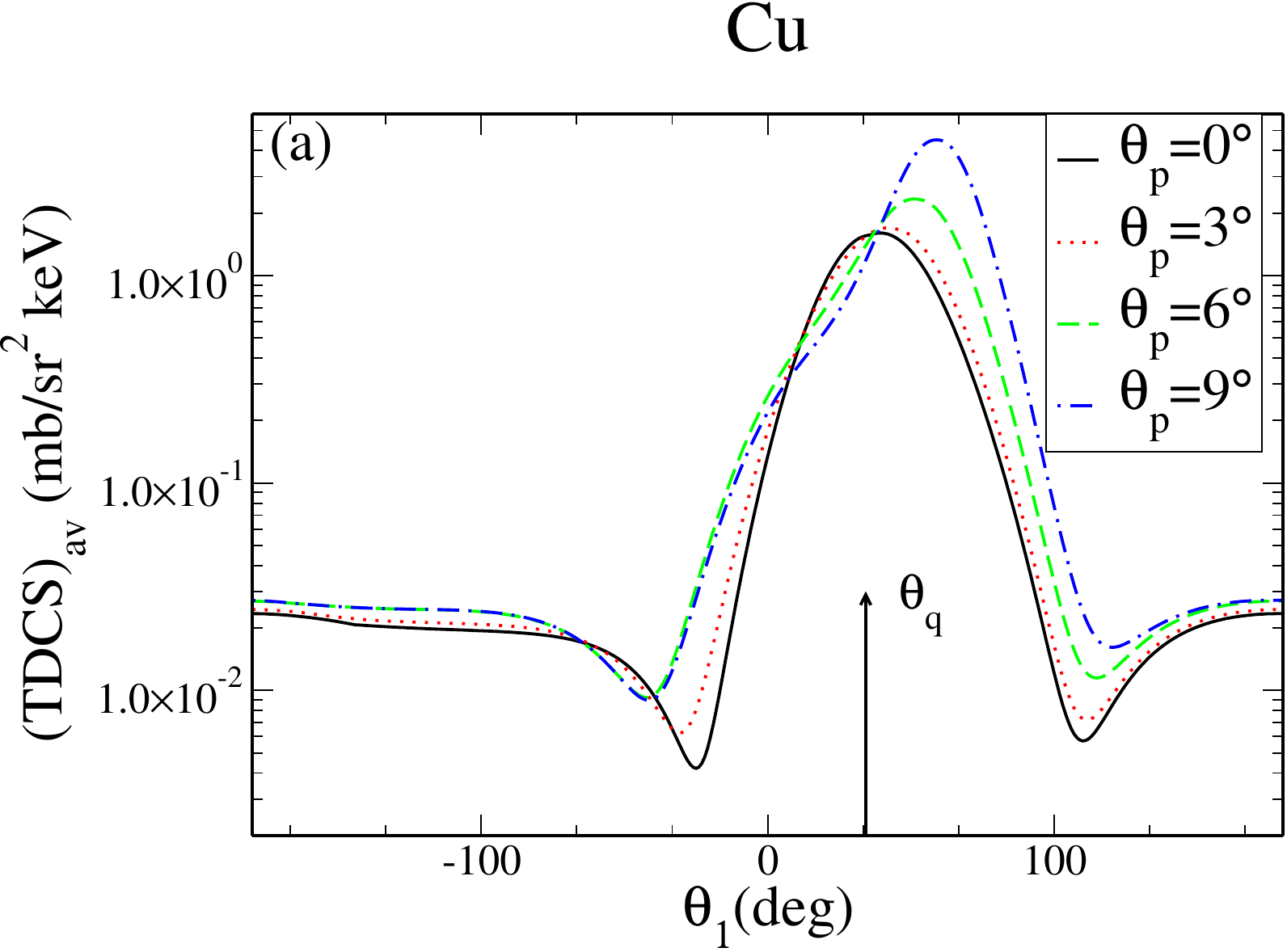}
\includegraphics[width = 0.8\columnwidth]{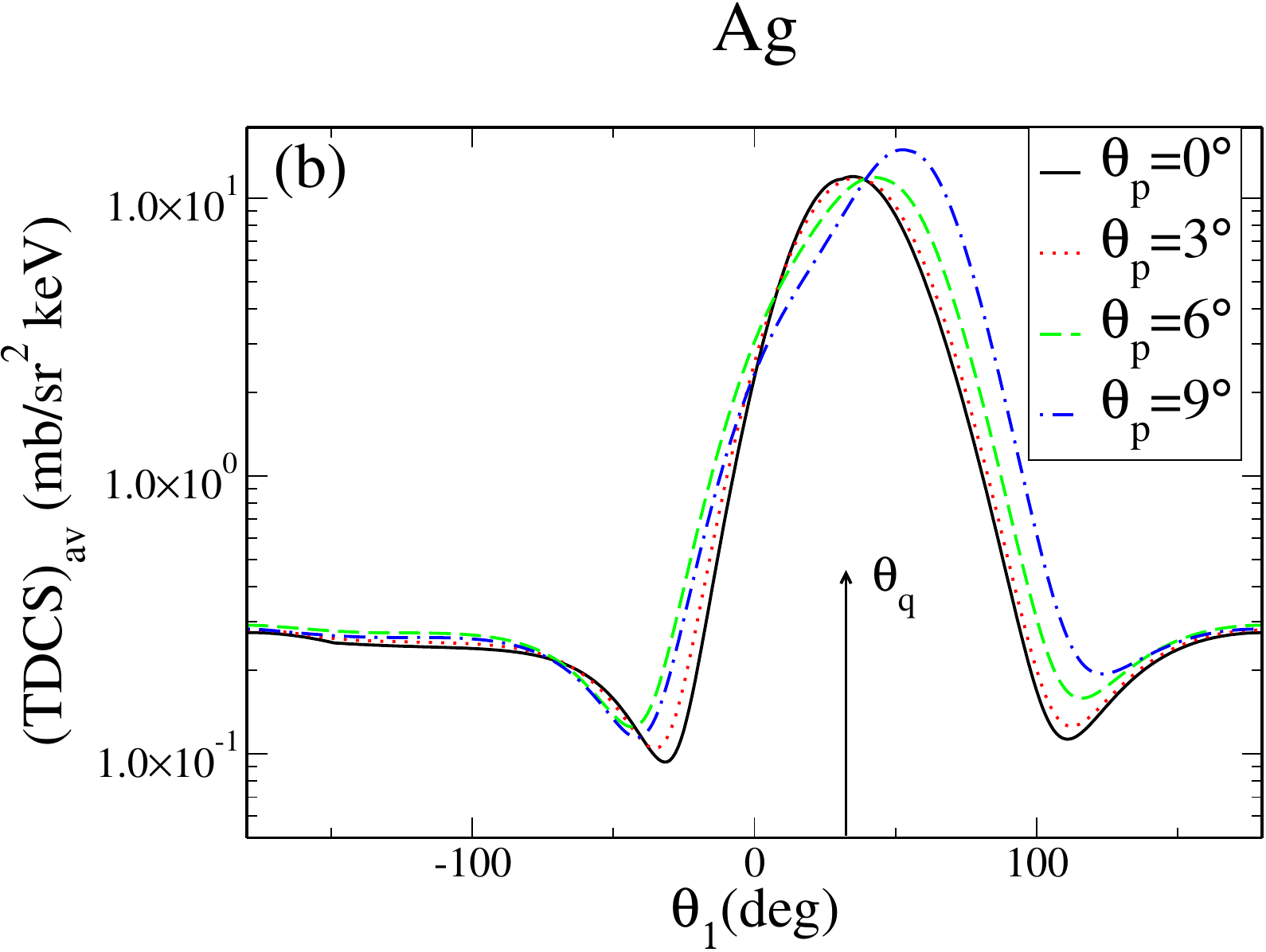}
\caption{TDCS averaged over impact parameter at $E_{i}=$300 keV with $\theta_{s} = 9^{\circ}$ and $10^{\circ}$ for Cu and Ag respectively with different $\theta_{p}$ as shown in the figure. Other kinematical variables are same as Fig.~\ref{fig5}.}\label{fig7}
\end{figure}

Having discussed \textcolor{black}{the} results of \textcolor{black}{the} TDCS for 300 keV, we \textcolor{black}{now} discuss the angular profile of TDCS for Ag target for given $l$'s for 500 keV. We also increase the scattering angle and hence the opening angle $(\theta_{s}=15^{\circ}$ and $\theta_{p}=15^{\circ})$ to investigate the effect of \textcolor{black}{larger} momentum transfer (larger scattering angle here) and the opening angle of the twisted beam on the angular profile of TDCS. The angular profiles of $(TDCS)_{T}$ for different values of $l$ are plotted in  Fig.~\ref{fig6}. As followed in Fig.~\ref{fig5}, we depict the various calculations of different $l$ with the same representative curves. For Ag case, we found that the binary peak for the plane wave is shifted by larger angles from the momentum transfer direction (see the solid curve in Fig.~\ref{fig6}). \textcolor{black}{When we gradually increase $l$ from 1 to 3, as found in the previous calculation, the binary peak in $(TDCS)_{T}$  shifts towards the forward direction. The binary peak split marginally for $l$=2 and prominently for $l$=3 leading to a two peak pattern in $(TDCS)_{T}$  in the binary region (see the dashed-dotted curve in Fig.~\ref{fig6})}. Further, we observe that the magnitude of $(TDCS)_{T}$ is more enhanced for $l$=1 when compared with the plane wave (see the solid and dotted curves of Fig.~\ref{fig6}.

\begin{figure}[htp]
\includegraphics[width = 0.8\columnwidth]{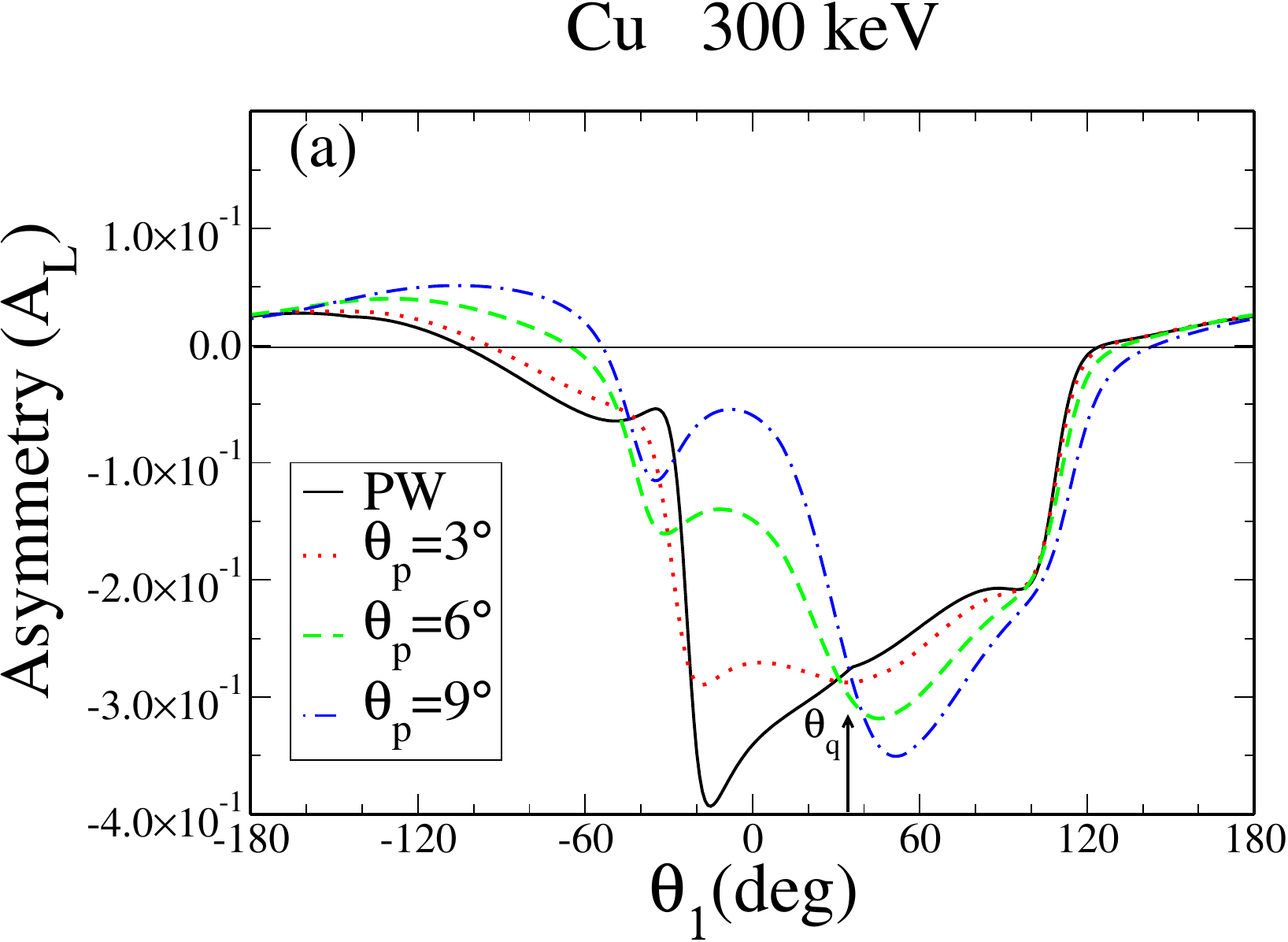}
\includegraphics[width = 0.8\columnwidth]{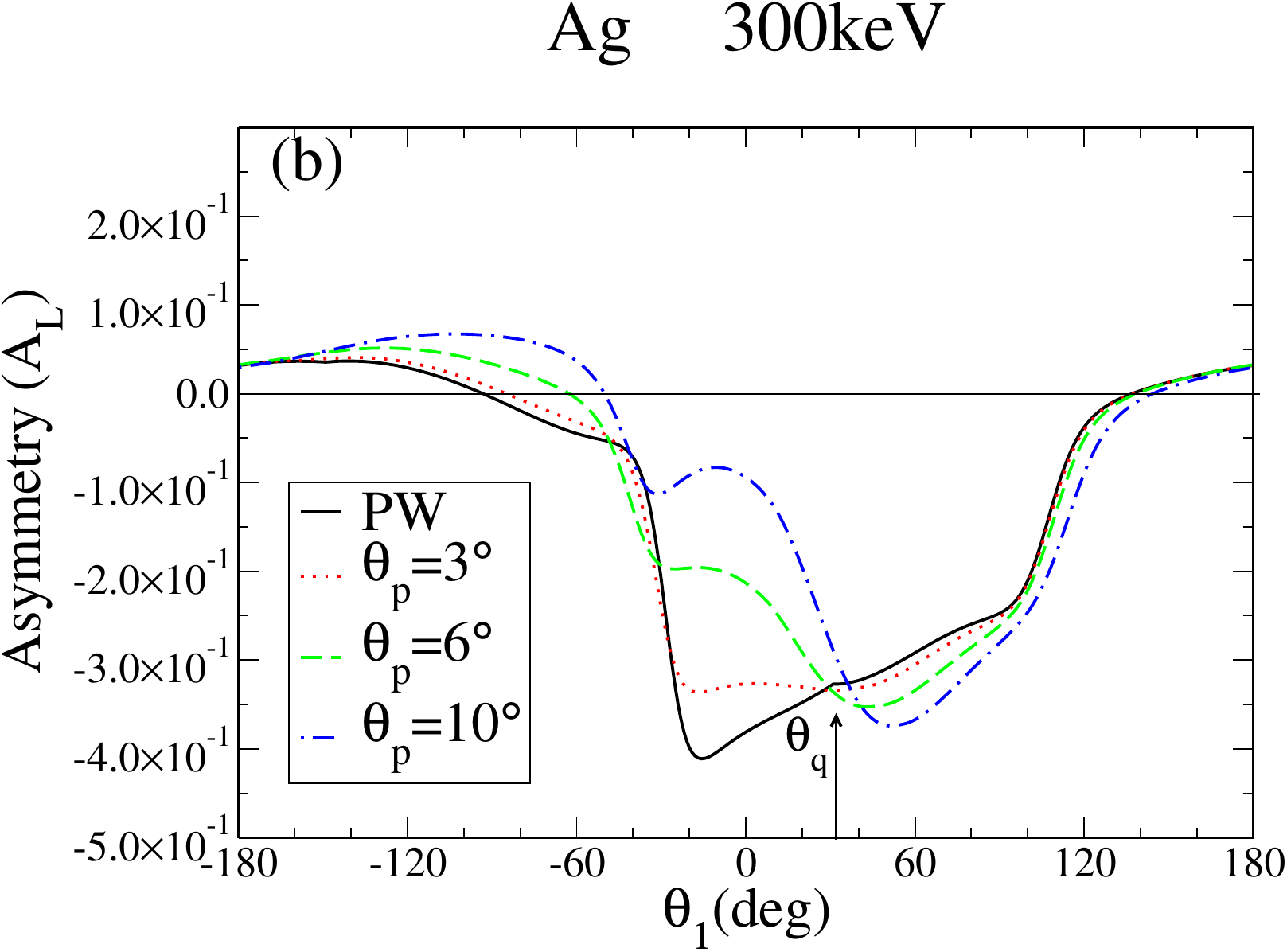}
\caption{Spin asymmetry $A_{L}$ in TDCS averaged over impact parameter ($(TDCS)_{av}$) as a function of the angle of the ejected electron for Cu and Ag (at $E_{i}=$300 keV) respectively. Other kinematical variables are same as Fig.~\ref{fig5}.}\label{fig8}
\end{figure}

We present the results of TDCS averaged over impact parameter ($(TDCS)_{av}$) for Cu and Ag targets in Fig.~\ref{fig7} \textcolor{black}{for an} incident energy $300$ keV in the coplanar asymmetry geometry. We use the same kinematics as used in Fig.~\ref{fig5} for the Cu and Ag. The advantage of \textcolor{black}{the} study of the angular profile of $(TDCS)_{av}$ is that it mimics the experimental realization of (e,2e) process on the foil target, like Cu and Ag, wherein there is a large collection of identical atoms in the transverse orientation of the target. We present the angular profile of $(TDCS)_{av}$ for Cu target in Fig. ~\ref{fig7}(a) and that for Ag target in Fig. ~\ref{fig7}(b). We keep $\theta_{s} = 9^{\circ}$ and $10^{\circ}$ for Cu and Ag respectively and vary $\theta_p$ as shown in Fig. ~\ref{fig7}(a) and ~\ref{fig7}(b). We represent the $(TDCS)_{av}$ for plane wave ($\theta_{p} =0^{\circ}$) by solid curve, $\theta_{p} = 3^{\circ}$ by dotted curve, $\theta_{p} =6^{\circ}$ by dashed curve and $\theta_{p} =9^{\circ}$ (for Fig.~\ref{fig7}(a)) or $\theta_{p} =10^{\circ}$ (for Fig. ~\ref{fig7}(b)) by dashed-dotted curve. From the angular profile of $(TDCS)_{av}$ for different $\theta_{p}$, we observe that the binary peak position shifts when we vary $\theta_{p}$, especially for the larger values of $\theta_{p}$. In  both the cases, the plane wave (solid line) and $\theta_{p}= 3^{\circ}$ (dotted curve) $(TDCS)_{av}$ calculations follow almost identical behaviour. However, for  other values of $\theta_{p}$, the angular profiles of $(TDCS)_{av}$ deviate significantly from that for plane wave case (see dashed and dashed-dotted curve and compare them with the solid curve). The binary peak shifts to a larger angle for larger  $\theta_p$. For example, the binary peak for  $\theta_{p} = 9^{\circ}$ and $\theta_{p} = 10^{\circ}$, used for Cu and Ag targets respectively, shifts to more than $15^{\circ}$ from the plane wave peak position (see dashed-dotted curve's binary peak positions). \textcolor{black}{Further}, we also observe that the magnitude of $(TDCS)_{av}$ for $\theta_{p} = \theta_{s}$ case is enhanced from that for the plane wave. This also validates the earlier finding of Serbo et al. (2015)~\cite{a25} wherein they observed enhanced differential cross section in the angular distribution when the scattering angle of the scattered electron equals the opening angle of the twisted electron. We would like to add here that Serbo et al.~\cite{a25} discussed the Rutherford like scattering of the incident electron from the Yukawa potential. On the other hand, we discuss here the (e,2e) process on atomic target which is different from the Rutherford like scattering. Despite of this, our calculation agrees with a similar conclusion as discussed above.

In the quantum mechanical complete experiment, we need to see the effect of spin of the impinging electron on the (e,2e) process. In the case of the polarized incident electron beam, the TDCS depends on the spin of the incident electron. This can be attributed to the spin orbit coupling of the electron when it is seen from the rest frame of the moving electron. In the rest frame of the moving electron, the intrinsic magnetic moment of incident electron due to its spin couples with the electromagnetic field of the atomic target. This leads to different types of coupling for $\lambda_{+}= \frac{1}{2}$ and $\lambda_{-}= -\frac{1}{2}$ spin states of the incident electron beam which should be reflected in the asymmetry $A_{L}$ of TDCS (see equation (\ref{eq18})). 

\textcolor{black}{Serbo et al. (2015) used a longitudinally polarized twisted electron beam with $\lambda = +{1/2}$ to investigate the polarization of the scattered electron in the Mott scattering on the macroscopic targets. In this communication, we investigate the spin asymmetry $A_{L}$ in the $(TDCS)_{av}$ for the macroscopic targets Cu and Ag.} \textcolor{black}{When we investigate the asymmetry $A_{L}$ of averaged over impact parameter TDCS ($(TDCS)_{av}$) for \textcolor{black}{the macroscopic target for} the twisted electron \textcolor{black}{beam}, we have to take into account the effects of opening angle $\theta_{p}$ on $A_{L}$ (as $(TDCS)_{av}$ doesn't depend on the TAM projection $m$). We plot spin asymmetry $A_{L}$ as a function of the ejection direction of the ejected electron for different $\theta_{p}$. We present the result of $A_{L}$ in Fig.~\ref{fig8} for Cu (frame (a)) and Ag \textcolor{black}{(frame (b))} at 300 keV. We keep $\theta_{p}=0^{\circ}$ (plane wave), $3^{\circ}$, $6^{\circ}$ and $9^{\circ}$ for the Cu target and $\theta_{p}=0^{\circ}$ (plane wave), $3^{\circ}$, $6^{\circ}$ and $10^{\circ}$ for the Ag target. We keep the same kinematics and follow the same representations of the curves as used in Fig.~\ref{fig7}. When we compare the spin asymmetry for the Cu and Ag targets in  Fig.~\ref{fig8}, we observe that for all the cases, the spin asymmetry varies with the ejection angle of the ejected electron. For all the cases of $\theta_{p}$, the spin asymmetries are mostly negative. In the forward region around $\theta_{1}= 0^{\circ}$, there is a maximum asymmetry observed in both the targets for the plane wave is used (see solid curves in Fig. ~\ref{fig8}). However, as $\theta_{p}$ increases, the spin asymmetry gradually decreases in the forward direction. For example, it is around -0.05 and -0.075 for the Cu and Ag targets respectively when $\theta_{p}$ is maximum ($\theta_{p}= 9^{\circ}$ and $10^{\circ}$ respectively for Cu and Ag targets) whereas for the plane wave case\textcolor{black}{,} it is around -0.4 for both the targets (see solid and dashed-dotted curves in Fig. ~\ref{fig8}). However, in the binary peak region, they follow the reverse trends resulting in maximum asymmetry for the highest $\theta_{p}$ (see solid and dashed-dotted curves in Fig. ~\ref{fig8}) in the binary peak region. They also follow similar trends in the recoil region around $\theta_{1}= 100^{\circ}$. In this region, the spin asymmetr\textcolor{black}{ies are} largely positive for all the calculations. \textcolor{black}{We end this section with a comment that} the spin asymmetry in the $(TDCS)_{av}$ depends on the opening angle of the twisted electron}.

\section{Conclusion}\label{sec4}

In this paper, we have studied the relativistic electron impact ionization of atomic targets by the twisted electron beam in coplanar asymmetric geometrical mode. We used a semi-relativistic Coulumb wave model for the computation of TDCS in the first Born approximation. A range of kinematical parameters was used in order to study the effect of various parameters of the twisted electron beam on the (e,2e) process. The spin asymmetry in TDCS averaged over impact parameter caused by the polarized incident electron beam is presented to \textcolor{black}{study} the effects of the twisted electron beam on (e,2e) process. We \textcolor{black}{also} studied TDCS for the charge density, sum of the charge and current density's part and total contributions\textcolor{black}{,} which includes the interference term of the matrix elements of the said terms. We observed significant dependence of TDCS and spin asymmetry on different parameters of the twisted electron beam and also on the atomic number $Z$.   

We would like to add that in \textcolor{black}{the} literature there are better models as compared to our FBA model for the plane wave (e,2e) process\textcolor{black}{.} \textcolor{black}{These models} can be explored for the twisted electron (e,2e) processes for further investigation on the effects of the twisted electron beam on it. Different kinematics, such as coplanar symmetric geometry, non-coplanar geometries, Bethe-Ridge kinematics, etc. can be considered. In the present model, the exchange between the scattered electron and the ejected electron can be considered for further study in \textcolor{black}{the} symmetric geometry mode. Further, one can explore a better relativistic wave function for the bound state. Although, at present, we don't have any experimental and theoretical results \textcolor{black}{of (e,2e) process} for the twisted electron beam at relativistic energy. Still, we are hopeful that this study may stimulate more studies in the field of electron impact ionization with the twisted electron beam, both at the theoretical as well as experimental levels.

\section*{Acknowledgement}
Authors acknowledge the Science and Engineering Research Board, Department of Science and Technology, Government of India, for funding the project through the Grant number EMR/2016/003125.

%%%%%%%%%%%%%%%%%%%%%%%%%%%%%%%%%%%%%%%%%%%%%%%%%%%%%%%%%%%%%%%%%%%%%%%%%%%%%%

%%%%%%%%%%%%%%%%%%%%%%%%%%%%%%%%%%%%%%%%%%%%%%%%%%%%%%%%%%%%%%%%%%%%%%%%%%%%%%
\bibliographystyle{apsrev4-1}
%\bibliography{ref_1}%

\end{document}